\documentclass[twocolumn,twocolappendix,tighten,appendixfloats]{aastex6}
\usepackage{natbib}
\usepackage{amssymb}
\usepackage{multirow}
\usepackage{graphicx}
\usepackage{tikz}
\usepackage{rotating}

\newcommand{\hi}{H {\small I}}
\newcommand{\hii}{H {\small II}}

\newcommand{\oiii}{[O {\small III}]}
\newcommand{\oii}{[O {\small II}]}

\newcommand{\hb}{H$\beta$}

\newcommand{\ha}{H$\alpha$}

\newcommand{\Jband}{$J$}
\newcommand{\Hband}{$H$}
\newcommand{\Kband}{$Ks$}
\makeatletter
\newcommand*{\rom}[1]{\expandafter\@slowromancap\romannumeral #1@}
\makeatother

\begin{document}
\submitted{Submitted: 26 June 2018}
\accepted{19 September 2018}

\title{Mapping Lyman Continuum Escape in Tololo 1247$-$232}


\author{Genoveva Micheva\altaffilmark{1,2}, M. S. Oey\altaffilmark{1},
  Ryan P. Keenan\altaffilmark{1,3}, Anne E. Jaskot\altaffilmark{4,5},
  and Bethan L. James\altaffilmark{6}} 



\altaffiltext{1}{University of Michigan, 311 West Hall, 1085 South University Ave, Ann Arbor, MI 48109-1107, USA}
\altaffiltext{2}{Leibniz Institut f\"ur Astrophysik, An der Sternwarte 16, D-14482 Potsdam, Germany }
\altaffiltext{3}{Steward Observatory, University of Arizona, 933 N. Cherry Ave, Tucson, AZ 85721, USA}
\altaffiltext{4}{Department of Astronomy, University of Massachusetts, Amherst, MA 01003, USA} 
\altaffiltext{5}{Hubble Fellow}
\altaffiltext{6}{Space Telescope Science Institute, 3700 San Martin Drive, Baltimore, MD 21218, USA}

\begin{abstract}
  Low redshift, spatially resolved Lyman continuum (LyC) emitters allow
us to clarify the processes for LyC escape from these starburst galaxies.
  We use {\it Hubble Space Telescope (HST)} WFC3 and ACS imaging of the
  confirmed low-redshift LyC emitter Tol 1247$-$232 to study the
  ionization structure of the gas and its relation to the ionizing star clusters. 
  We perform ionization parameter mapping (IPM) using [O
  \small{III}]$\lambda\lambda4959,5007$ and [O
  \small{II}]$\lambda3727$\normalsize\ imaging as the high- and
  low-ionization tracers, revealing broad, large-scale, optically thin
  regions originating from the center, and reaching the outskirts of
  the galaxy, consistent with LyC escape.  We carry out stellar
  population synthesis modeling of the 26 brightest clusters using
  our {\it HST}  photometry.  Combining these data with the nebular
  photometry, we find a global LyC escape fraction of $f_{\rm esc} =
  0.12$, with uncertainties also consistent with zero escape and with all
  measured $f_{\rm esc}$ values for this galaxy. Our analysis 
  suggests that, similar to other candidate LyC emitters, a two-stage 
  starburst has taken place in this galaxy, with a $12$ Myr old,
  massive, central cluster likely having pre-cleared regions in and
  around the center, and the second generation of $2-4$ Myr old
clusters dominating the current ionization, including some escape from the galaxy.
\end{abstract}


\section{Introduction} \protect\label{sec:intro}
Escape of Lyman continuum (LyC) radiation from local star-forming
galaxies (SFGs) has recently been detected from a handful of local
galaxies at redshifts $0.02\leq z\lesssim0.33$
\citep[][]{Bergvall2006,
  Leitet2011,Leitet2013,Leitherer2016,Borthakur2014,Izotov2016a,Izotov2016b,Chisholm2017,Izotov2018}. The
emitting galaxies comprise a sample of four blue compact galaxies
\citep[BCGs; e.g.,][]{Thuan1981}, one Lyman Break analog
\citep[LBAs;][]{Heckman2005,Overzier2009}, and six Green Peas
\citep[GPs;][]{Cardamone2009}.  The confirmed low-redshift LyC
emitting GPs \citep{Izotov2016a,Izotov2016b,Izotov2018} have 
escape fractions in the range $6\mbox{-}46$\%, but being at
redshifts $z\sim0.3$ they remain unresolved even in Hubble Space
Telescope ($HST$) images. These spatially unresolved galaxies do not
allow for a detailed investigation of the mechanisms behind the
leakage of LyC. 

The four LyC-emitting BCGs Haro 11, Tol 1247$-$232, Mrk 54, and
Tol 0440--381 \citep{Bergvall2006,Leitet2013,Leitherer2016,
  Chisholm2017}, have lower escape fractions of a few percent,
but are much closer, at $0.02\leq z\leq0.048$. At such redshifts, $HST$
resolution is able to reveal the morphology and ionization structure of the
interstellar medium (ISM) and allows identification of
the star-forming regions responsible for LyC escape. Hence, these galaxies
offer the opportunity to clarify the processes behind the observed LyC
leakage. 

Tol 1247$-$232 is at a luminosity distance of $213$ Mpc ($z=0.048$)
and hence allows for such a spatially resolved investigation into the
morphology of the ionized gas and relation to super star clusters
(SSCs).  The escape fraction of Tol 1247$-$232 has
been measured four times, with two different instruments, being
consistently non-zero in all measurements
\citep{Leitet2013,Leitherer2016,Puschnig2017,Chisholm2017}. This galaxy
therefore offers a spatially resolved view of the
ionization structure and morphology of a confirmed LyC emitter (LCE). 

The technique of ionization parameter mapping (IPM) is well suited
for such an investigation. IPM was first applied to \hii\ regions in
the Small (SMC) and Large Magellanic Clouds \citep[LMC;
][]{Pellegrini2012}, to identify optically thin \hii\ regions and
trace the escape of LyC photons from star-forming regions into the 
interstellar medium.  Further applications of this technique
have detected large-scale optically thin regions in NCG 5253
\citep{Zastrow2011}, NGC 3125 \citep{Zastrow2013}, and Haro 11
\citep{Keenan2017}, consistent with LyC escape into the
circumgalactic medium and beyond.

Of these three galaxies, Haro 11 is a confirmed LCE
at $z=0.02$ \citep{Bergvall2006,Leitet2011}.  IPM analysis of this object
\citep{Keenan2017} reveals some surprising and important results.
Whereas Knot~C has been assumed to be the LyC source because of its 
strong Ly$\alpha$ emission, IPM indicates that this region is
optically thick in the LyC, and instead, Knot A is strongly indicated as the
source of the leakage.  IPM also suggests that Knot B may
be optically thin as well  \citep{Keenan2017}, and it may host a low-luminosity
active galactic nucleus \citep{Prestwich2015}, 
which is also conducive to escape of ionizing photons.  
It is therefore unclear which of the three bright knots in Haro 11 is
responsible for the observed escape of LyC flux, and in particular,
what the relationship is between LyC emission and Ly$\alpha$ emission.

In this paper, we explore the LyC radiative transfer in Tol
1247$-$232.  Here, IPM paints a much clearer picture, easily revealing the 
dominant source of the ionizing radiation.  In Section \ref{sec:data}
we present the broad- and narrowband {\it HST} imaging data used in
the analysis, describe the procedure of the continuum subtraction, and
the IPM technique.  Section \ref{sec:phot} presents the stellar population
analysis of all bright knots (``clusters'') inside the galaxy, and on the
galaxy as a whole.  We estimate the global LyC escape fraction in Section
\ref{sec:discussion}, discuss the implications of our analysis in Section
\ref{sec:discuss2}, and present our conclusions in Section
\ref{sec:conclude}. The appendix provides a comparison between different
methods for continuum subtraction in Section \ref{sec:mu_appendix}, and
supplementary data from the modeling of the spectral energy distribution of
the clusters in Section \ref{sec:seds_pdfs_appendix}. Throughout this paper,
we assume $\Lambda$CDM cosmology with $\mathrm{H_0}=70$ km s${}^{-1}$
Mpc${}^{-1}$, $\Omega_{\rm M}=0.3$, and $\Omega_{\rm \Lambda}=0.7$, and hence
a luminosity distance to Tol 1247$-$232 of ${D_L=213.1}$ Mpc at $z=0.048$.
All photometry is in the AB magnitude system. 

\section{Nebular Analysis}
\begin{figure}
  \gridline{
    \fig{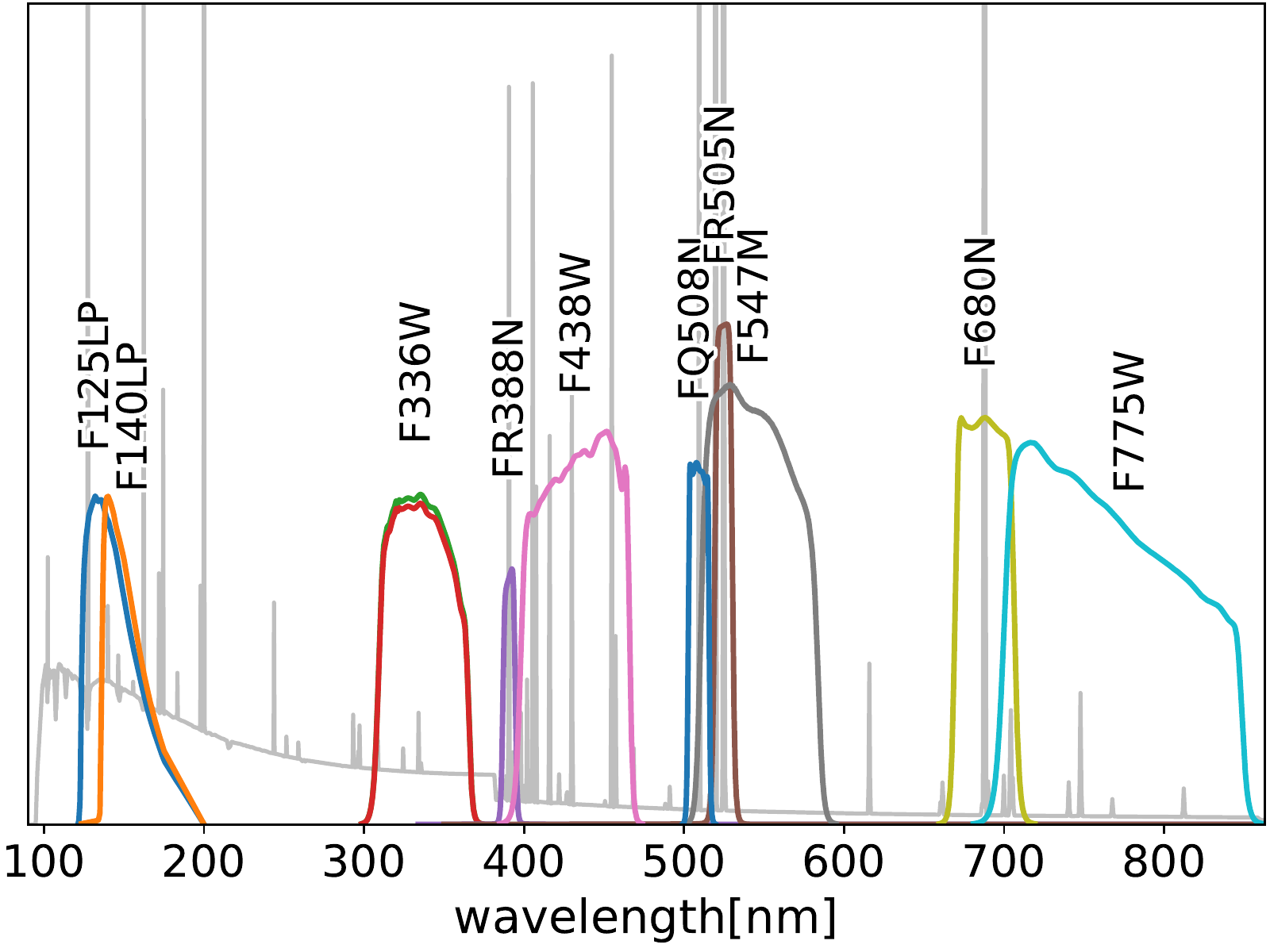}{0.45\textwidth}{}
  }
\caption{All $HST$ filters utilized in this work, with an example
  galaxy SED at the redshift of Tol 1247$-$232, $z=0.048$.  Note that
  F336W and F438W bracket the Balmer break at this
  redshift.\protect\label{fig:filters}}   
\end{figure}

\protect\label{sec:data}
We obtained {\it HST} WFC3 and ACS imaging in
broadband and narrowband filters (P.I.  Oey, PID 13702) that sample
the $U$ (F336W) and $V$ bands (F547M) on the WFC3, and [O {\small
    II}]$\lambda3727$ (FR388N) and [O {\small
    III}]$\lambda\lambda4959,5007$ (FR505N) on the ACS.  Additional
WFC3 data (P.I.  \"Ostlin\footnote{Spelled Oestlin in the HST proposal archive}, PID 13027) exist for this galaxy in $B$
(F438W), $R$ (F775W), \ha\ (F680N), \hb\ (FQ508N), and far ultraviolet
(FUV) F125LP and F140LP, which we used to complement the coverage of
our target.  The exposure times were $1208$ s (F125LP and F140LP), $1030$ s (F336W, P.I. Oey), $900$ s (F336W, P.I. \"Ostlin), 
$2710$ s (FR338N), $732$ s (F438W), $2355$ s (FR505N), $1340$ s (FQ508N), $1095$ s (F547M), $740$ s (F680N), and $600$ s (F775W).
We resampled all images to the larger pixel scale of the
ACS of $0.05$ arcsec.  The filters are illustrated in
Figure~\ref{fig:filters}.  Unless otherwise stated, hereafter we will
refer to [O {\small II}]$\lambda3727$ and [O {\small
    III}]$\lambda\lambda4959,5007$ as simply \oii\ and \oiii,
respectively. We note that the F125LP filter contains the Ly$\alpha$
  line, which cannot be modeled by standard population synthesis tools. The
  filter is broad, however, and is not dominated by line emission, so the
  line flux is not significant in our case.  

\begin{figure*}
\gridline{\fig{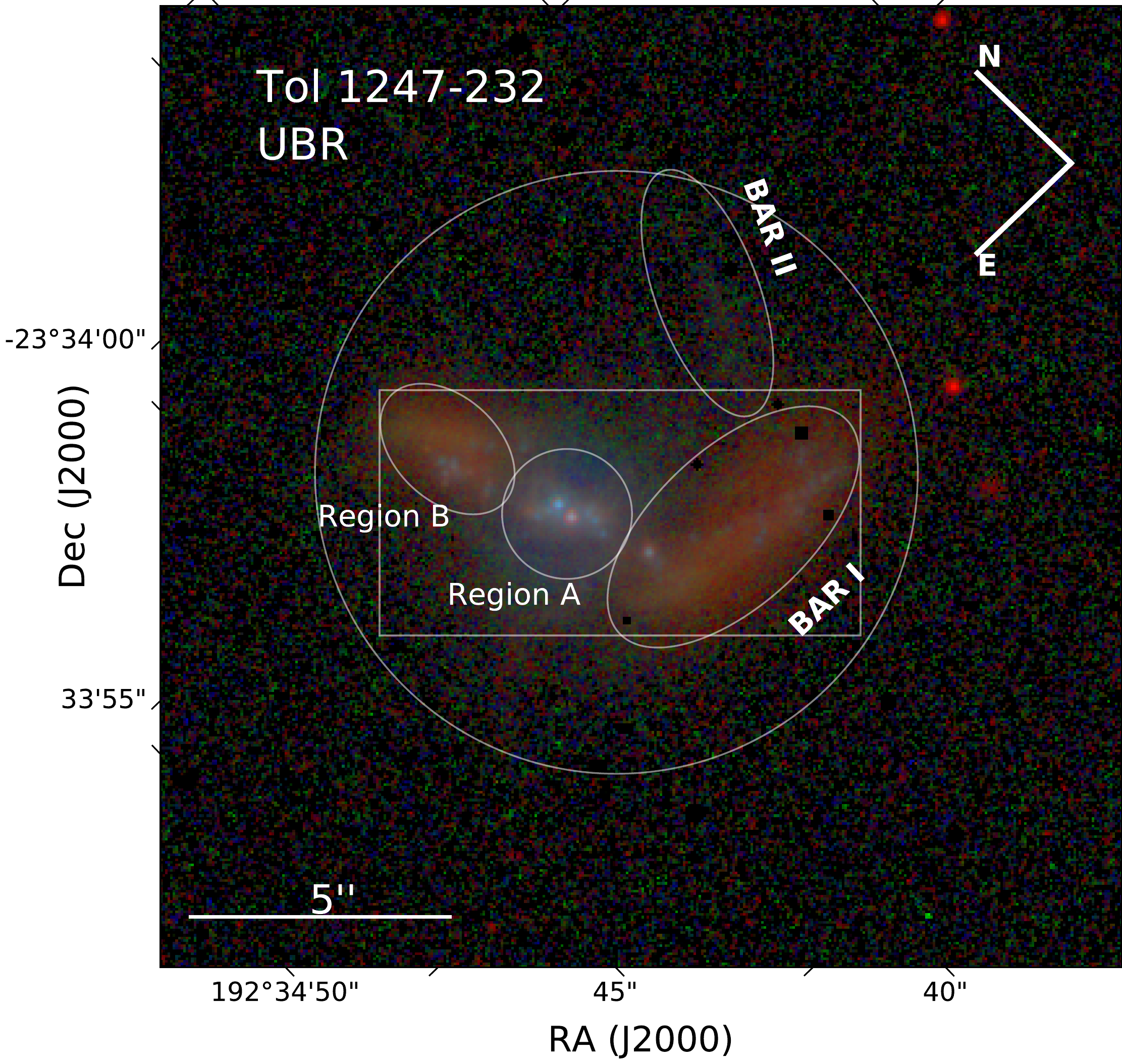}{0.3\textwidth}{(a)}
          \fig{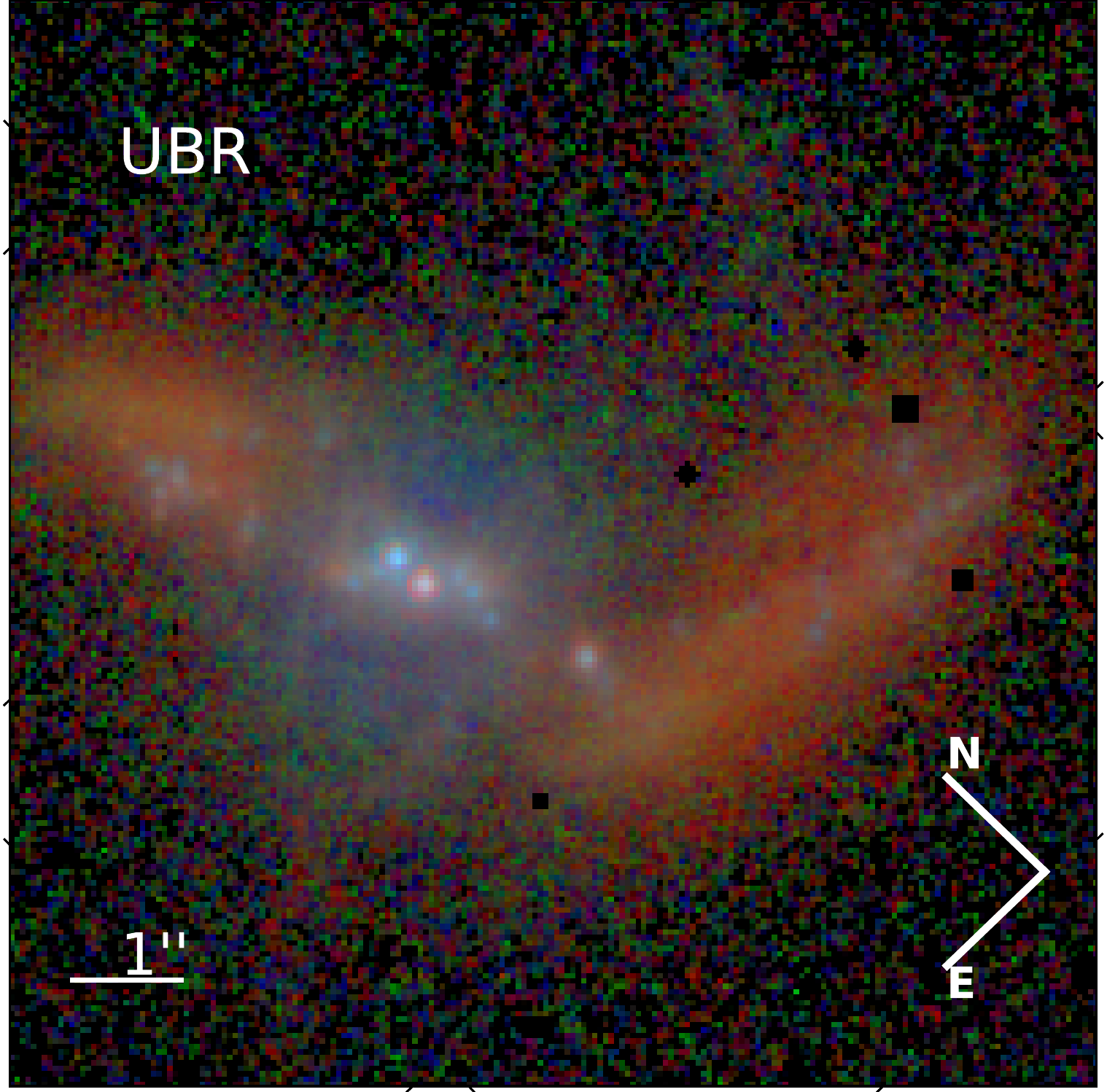}{0.3\textwidth}{(b)}
          \fig{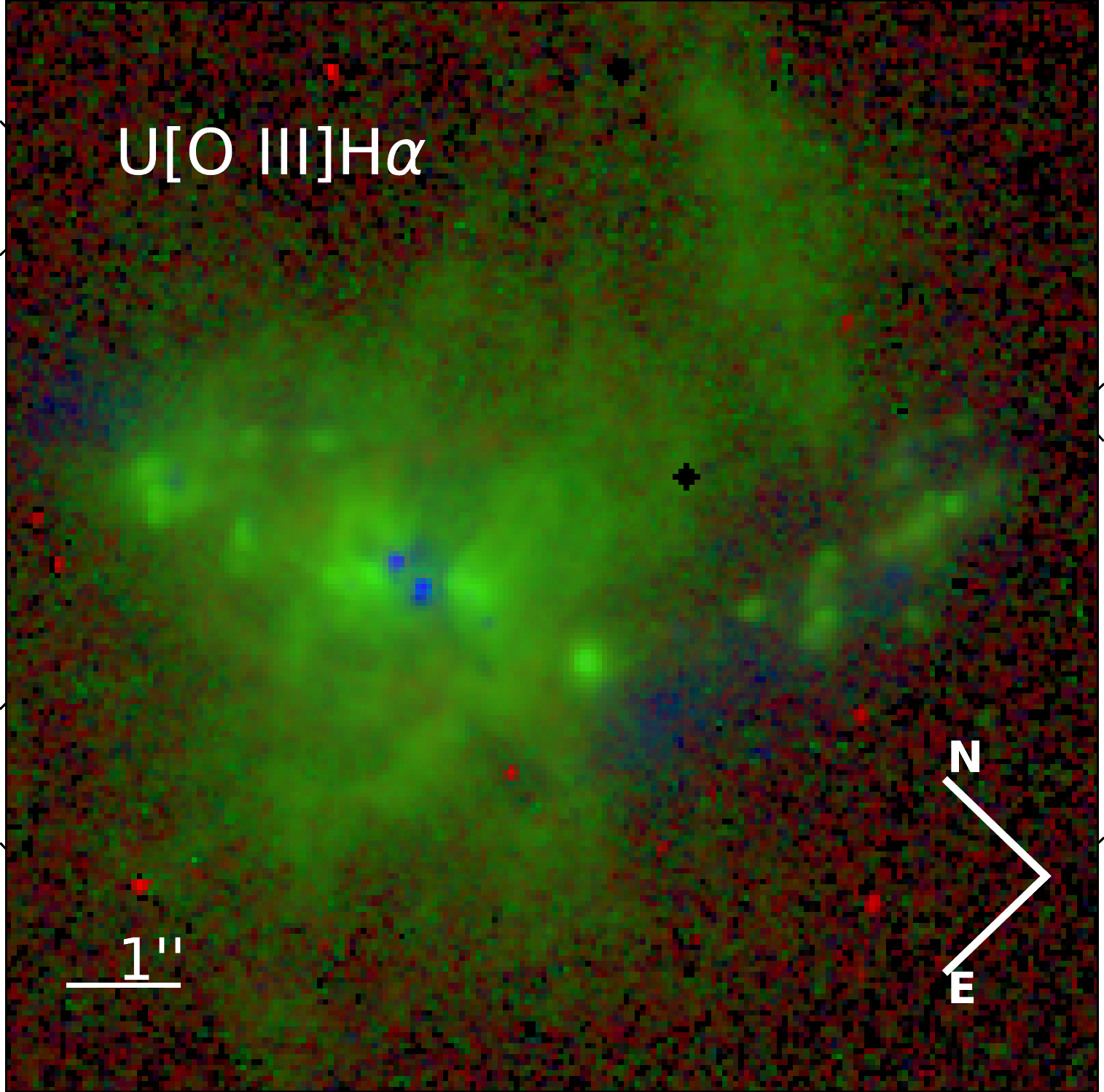}{0.3\textwidth}{(c)}
          }
\caption{(a) A false color image of Tol 1247$-$232 with broadband
  F336W, F438W, and F775W as the blue, green, and red channels,
  respectively.  The central Region A, coincident with the size and
  location of the $HST$/COS aperture in \citet{Puschnig2017}, is
  indicated with a small white circle.  Region B to the north of A is
  indicated by a white ellipse.  Bar I and Bar II to the southwest
  and west of A, respectively, are also marked by ellipses.  The white
  rectangle shows the area used to estimate the continuum scaling
  factor $\mu$ in Section \ref{sec:mu}.  The large white circle
  indicates the aperture used for photometry on the total galaxy area.
  (b) Zoom of the galaxy region.  Blue knots are visible not only in
  Region A, but also in B and Bar I.  (c) Zoom of the galaxy with
  broadband F336W, and continuum-subtracted \oiii, and \ha\ images as
  the blue, green, and red channels, respectively.  Bar II is clearly
  visible here, and appears to be emission
  line-dominated. At the distance of Tol 1247$-$232, $1$ arcsec corresponds to
  0.94 kpc.\protect\label{fig:rgb}}   
\end{figure*}

To calibrate  data, we used {\tt pysynphot} with the observing mode
given from the PHOTMODE header keyword, and assuming an infinite
aperture.  In practice, the aperture radius was set to $2$ arcsec for
the narrowband data, and $6$ arcsec for the broadband data.  All of
our photometry therefore is measured relative to an infinite aperture
zeropoint.  To facilitate discussion of different galaxy regions, we introduce a
naming convention, defined in Figure \ref{fig:rgb}a.   

\subsection{Continuum subtraction}\protect\label{sec:mu}
The four narrowband filters target the named \oii, \hb, \oiii, and \ha\
lines.  To isolate the line flux in each filter one must remove the
contribution of the underlying continuum.  
We use F336W, F438W, F547M, and F775W, respectively, as off-line continuum
filters, and estimate the line flux as $f_{\rm on line}-\mu f_{\rm off
  line}$.  %

There are several methods for obtaining the scaling factor $\mu$, producing either spatially
resolved values \citep[e.g.,][]{Hayes2009, James2016}, or a single effective
$\mu$ \citep[e.g.,][]{Boker1999,Kennicutt2008, Hong2014,
  Keenan2017}.  The advantage of the former is that it accounts for
spatial variations of the scaling factor, arising from its dependence
on the color of the underlying stellar population, which varies from
place to place.  Figure
\ref{fig:rgb}a shows that there are indeed color gradients evident in
Tol 1247$-$232, with the central Region A having blue colors hinting
at a young population, and redder, older populations to the north in
Region B and to the south-west in Bar I.  Figure \ref{fig:rgb}b is a
zoom of the galaxy, showing smaller-scale variations in the stellar population.
\citet{Puschnig2017} account for the presence of color gradients by
using spatially resolved SED fitting to estimate continuum
subtraction in their analysis of the distribution of Ly$\alpha$
emission in Tol 1247$-$232. However, this
method is highly model-dependent, and determining the continuum scale factor
is fraught with uncertainties, as we demonstrate in Section \ref{sec:seds}.  

We therefore use the mode method of \citet{Keenan2017} to estimate a
single scaling factor $\mu$ across the entire galaxy. This method
is based on evaluating the mode of the pixel histogram of the
continuum-subtracted image 
as a function of the scaling factor $\mu$. For each of the four
emission lines, the mode is computed over an area covering most of the
galaxy, indicated by the white rectangle in Figure \ref{fig:rgb}a. This area
does not include Bar II.  However, as can be seen from the 
superposition of Figures \ref{fig:rgb}b and \ref{fig:rgb}c, Bar II
contains almost no visible stellar emission, being strongly dominated by
nebular emission.  Including Bar II would therefore only contribute
background-dominated noise in determining the continuum scale factor,
and we have therefore omitted Bar II from the region used to estimate
$\mu$. Further, we note that the scaling factor one obtains from the white
rectangle in Figure \ref{fig:rgb}a and from only Region A, agree to $90\%$. 

\begin{figure*}
  \plotone{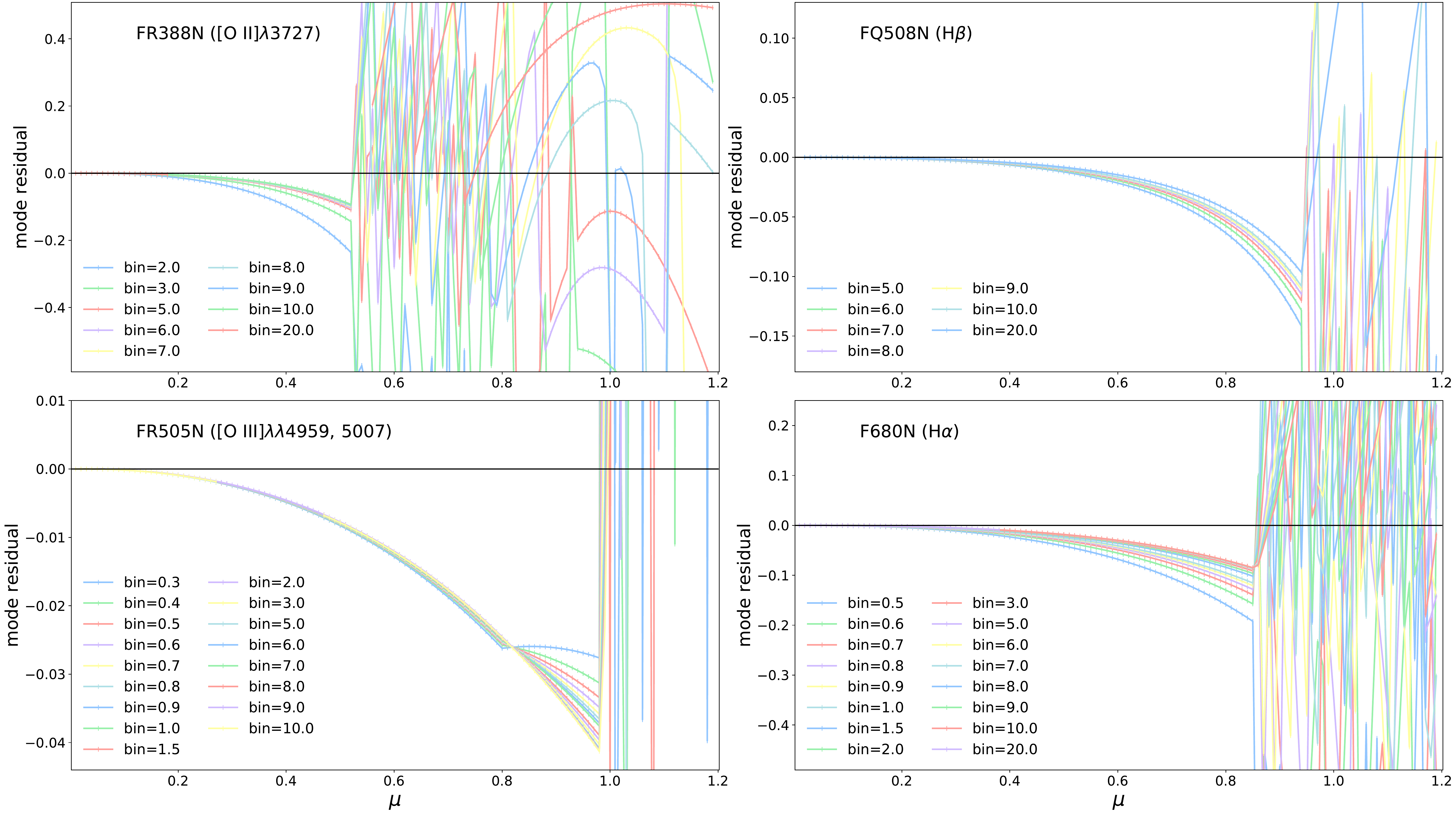}
\caption{
  Mode method for continuum subtraction \citep{Keenan2017}: Scaling factor $\mu$ as a function of
  the mode.  For convenience, on the $y$-axis, we plot the residual between
  the mode and a linear fit to the first few
  mode points. This presentation better highlights the breaks at
  optimal $\mu$. For all displayed bin sizes, the mode function shows
  a strong break at about the same $\mu$ value. In each
  panel this break represents the optimal scaling factors, which are
  $0.52$, $0.94$, $0.80$, and $0.84$ for \oii, \hb, \oiii, 
  and \ha, respectively.\protect\label{fig:scalingfactor}}   
\end{figure*}

Figure \ref{fig:scalingfactor} shows the break in the mode versus
$\mu$ function, indicating the location of the optimal $\mu$ for each
filter. This break indicates the transition from undersubtraction to
oversubtraction of the continuum \citep[see][]{Keenan2017}.  The observed breaks indicate
$\mu=0.52,\,0.94,\,0.80$, and $0.84$ for \oii, \hb, \oiii, and \ha,
respectively.  Since the mode depends on the bin size, we have
explored all bin sizes from $0.1$ to $20\sigma$ in steps of
$0.1\sigma$, where $\sigma$ is the standard deviation of the pixel fluxes
inside the white rectangle in Figure~\ref{fig:rgb}a. Beyond a certain bin
size, the behavior of the mode as a function of $\mu$ converges to produce a
break at the same position for all larger bin sizes.  In the figure we show
the mode function for different bin sizes, starting from the smallest bin
beyond which all bin sizes produce a break at the same $\mu$. We also show a
few larger bins to demonstrate the robustness of that convergence. The on-line
and off-line fluxes are corrected for Galactic extinction using the \citet{Schlafly2011} reddening curve 
before the subtraction. Due to line contamination in the
  F547M continuum filter, our \oiii/\oii\ ratios are lower limits, as
  discussed in Appendix \ref{sec:linecont}.

%

In Appendix~\ref{sec:mu_appendix}, we compare the mode method to
other methods for determining the global scale factor.  The agreement between the
methods is good, and we conclude that the mode method is a relatively
straightforward and efficient way to obtain the scaling factors.  One clear advantage
over other methods is that the break in the mode function, and hence
the value of $\mu$, is unambiguous and easy to identify.  In what
follows, we use the continuum-subtracted images produced with the mode
method.  

\begin{figure*}
  \gridline{
    \leftfig{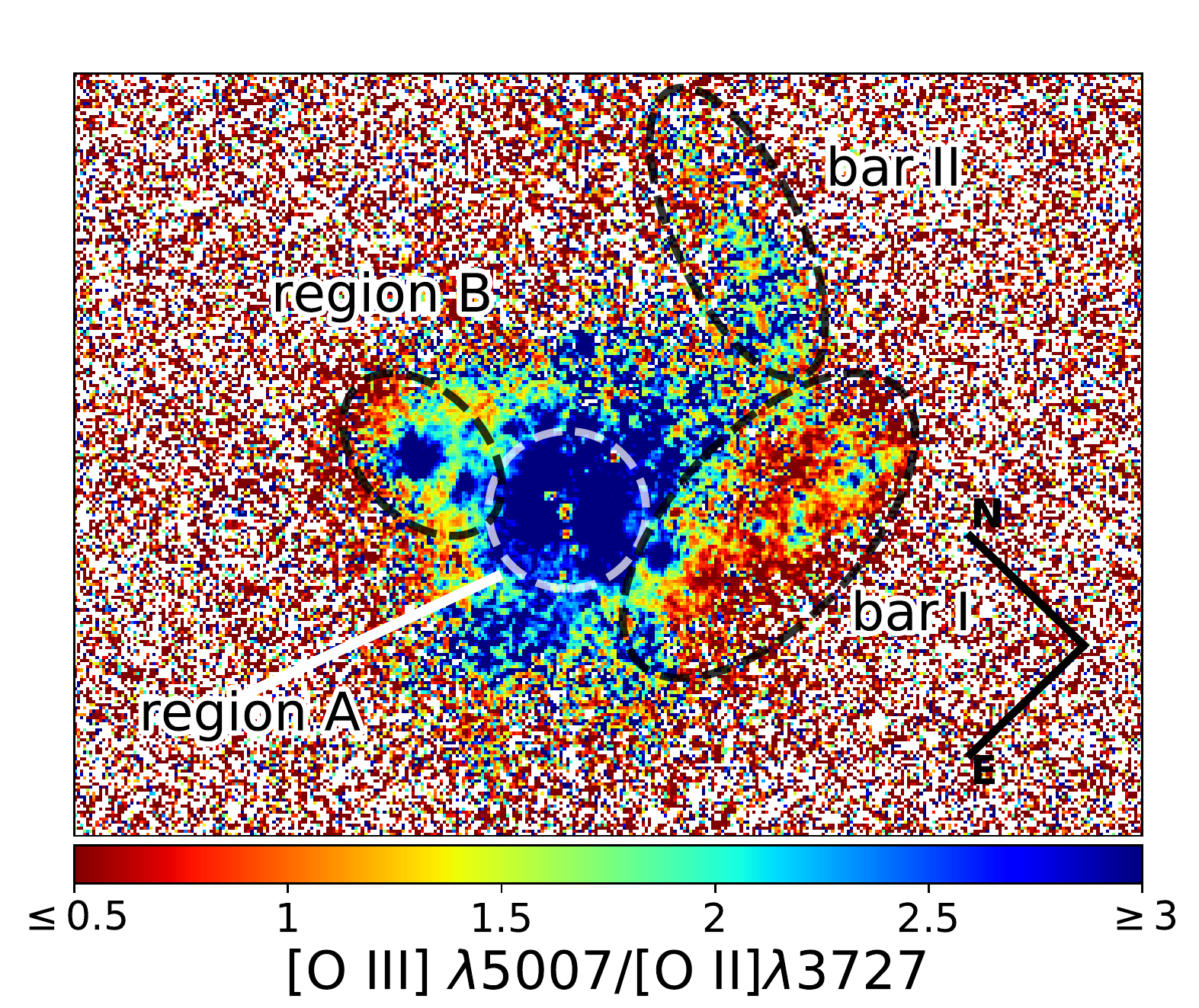}{0.5\textwidth}{(a)}
    \rightfig{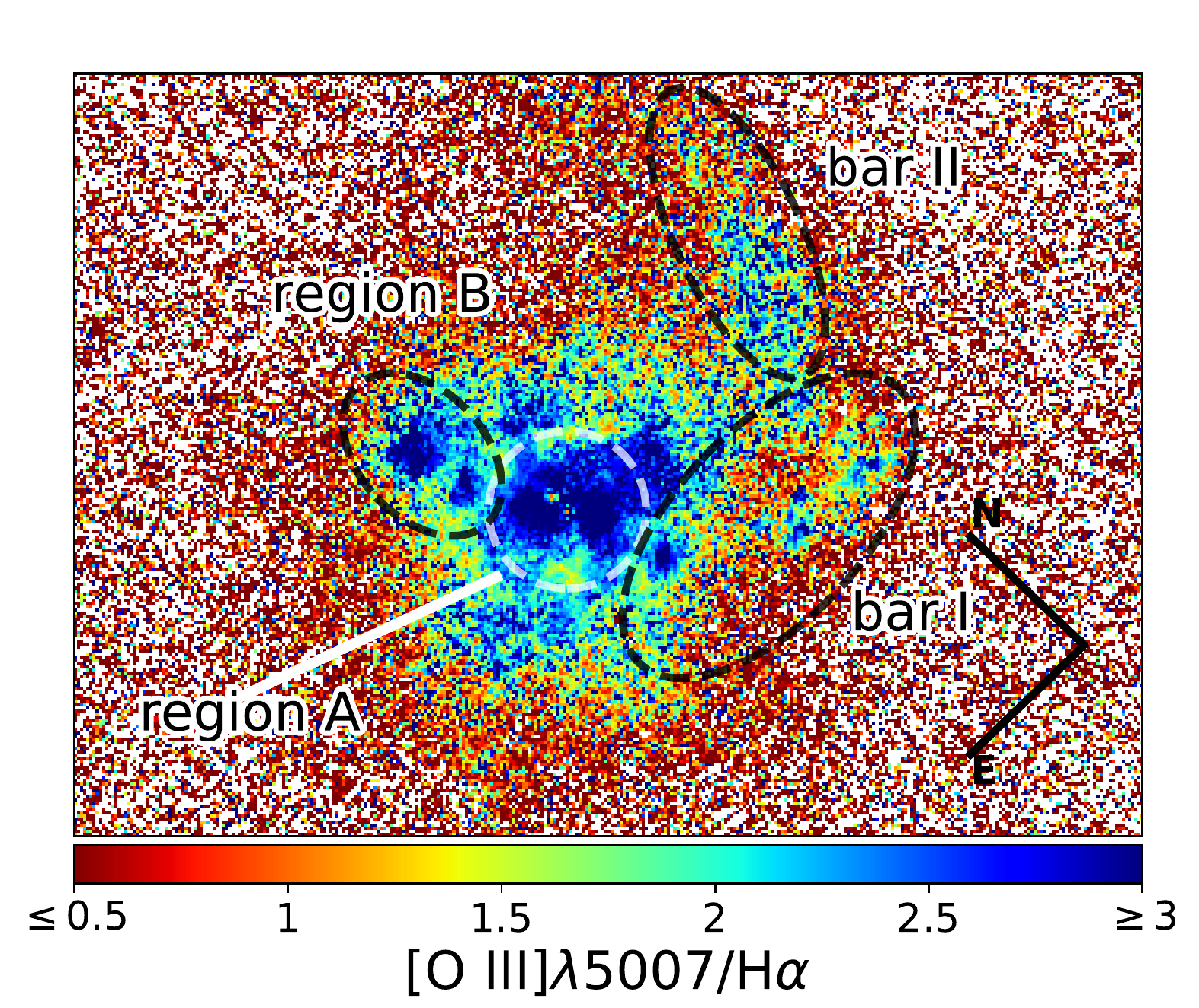}{0.5\textwidth}{(b)}
  }
  \gridline{
    \leftfig{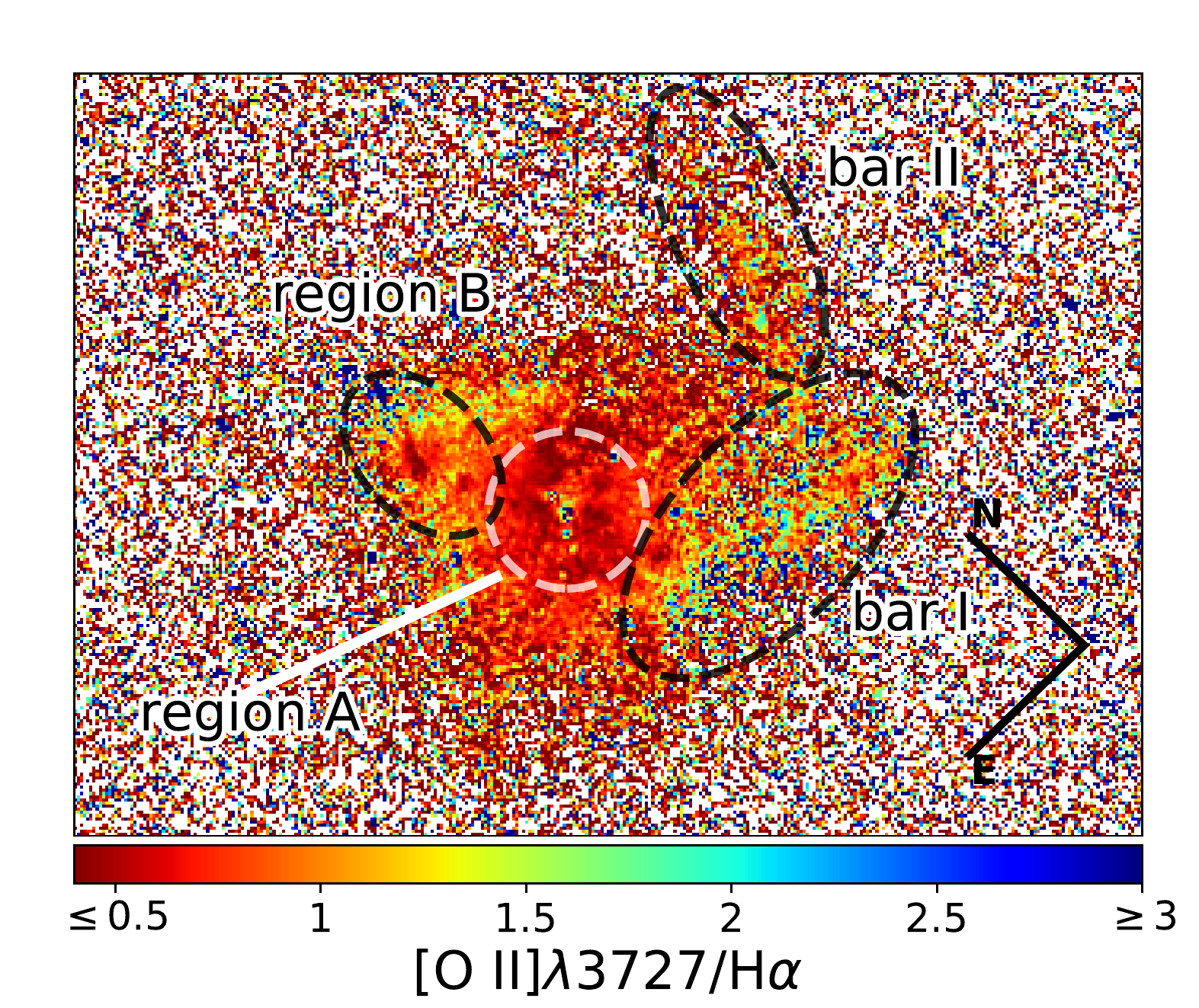}{0.5\textwidth}{(c)}
    \rightfig{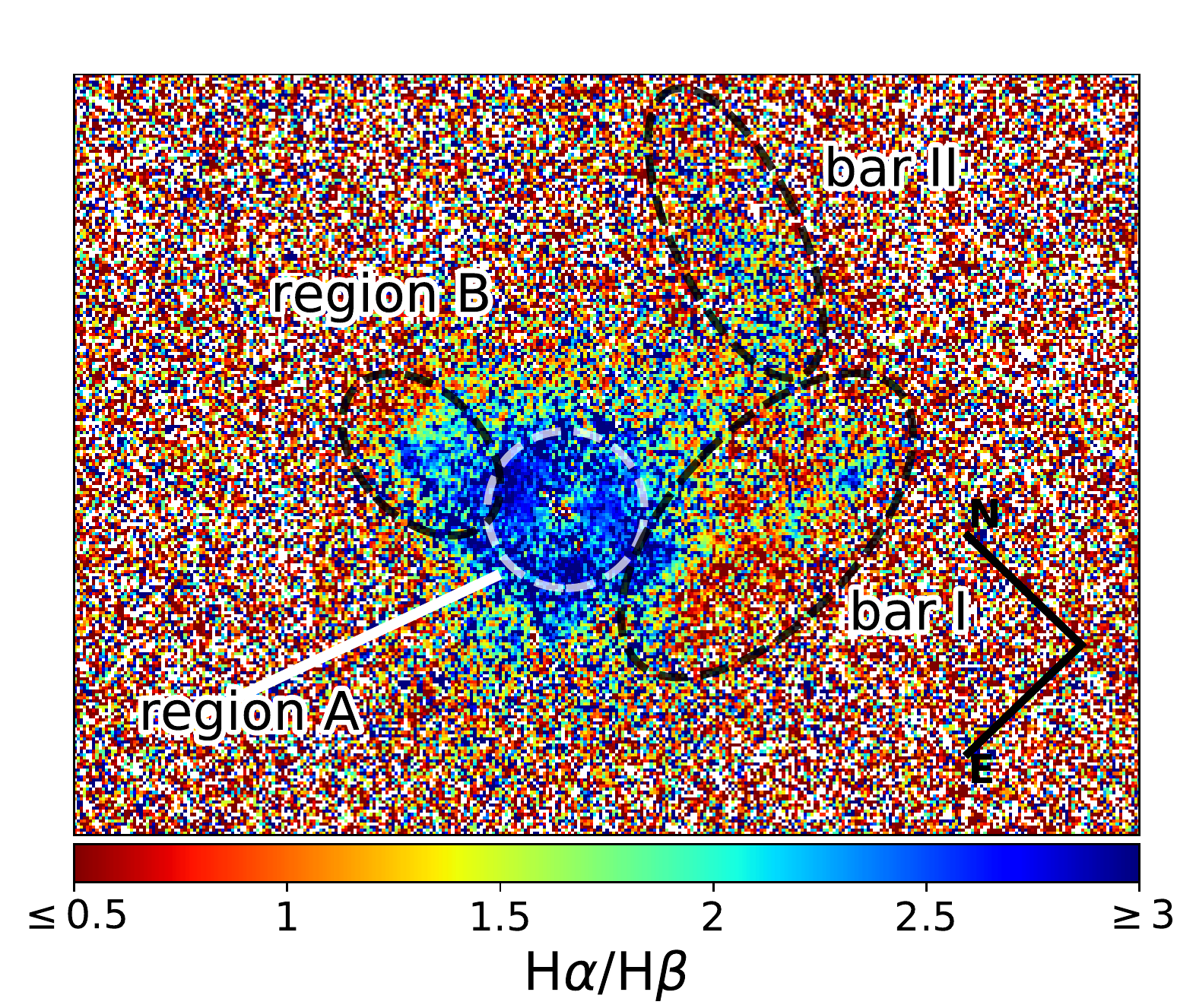}{0.5\textwidth}{(d)}
    }
\caption{Ionization parameter mapping.  Boundaries of Regions A (white dashed
  circle), B, Bar I and Bar II (black dashed ellipses), defined in
  Figure \ref{fig:rgb}a, are overplotted for orientation.  Panels (a),
  (b), (c) and (d) show the ratios of \oiii/\oii,
  \oiii/\ha, \oii/\ha, and \ha/\hb/ respectively.  All ratios are based on
  continuum-subtracted images as described in Section~\ref{sec:mu}.
  Region A has a diameter of $2.5$ arcsec, corresponding to $2.3$ kpc. \protect\label{fig:ipm}}   
\end{figure*}

\subsection{Ionization parameter mapping}\protect\label{sec:ipm}
The technique of ionization parameter mapping (IPM) spatially explores
the ionization structure of the interstellar medium
\citep[e.g.,][]{Pellegrini2012,Zastrow2011,Zastrow2013,Keenan2017}.
Regions with high ionization can be traced via, e.g., O$^{+2}$ or
S$^{+2}$ emission, which require photons with energies significantly
higher than what is needed to ionize hydrogen.  The detection of lines
from these species therefore indicates that hydrogen is predominantly
ionized in such regions.  Further away from the source of ionizing
photons, the ionization state of hydrogen ordinarily transitions to
neutral, for optically thick conditions.
In this transition region, low-ionization species dominate and can
be traced via, e.g., O$^{+1}$ or S$^{+1}$ lines.  While other species
can be used as tracers, a practical constraint is that the
emission lines be strong enough to be easily detected. 

We use our continuum-subtracted \oiii\ and \oii\ imaging to
serve as the high and low ionization tracers, respectively. The
\oiii/\oii\ ratio, often strong in young starbursting regions, is
a good proxy of the ionization parameter $U$
\citep[e.g.,][]{Jaskot2013}. Figure \ref{fig:ipm}a shows the \oiii/\oii\
ratio. As discussed in Appendix \ref{sec:linecont}, the \oiii/\oii\ values are
lower limits, due to the oversubtraction of the continuum in the FR505N
filter. High values of this ratio (coded blue) are 
indicated throughout the central region of the galaxy, including along the
minor axis both to the east and west of the center, clearly reaching
the outskirts of the galaxy. These ionized regions are broad,
extending $\sim3$ kpc from the center of Region A in both directions. 
Region A itself appears completely ionized, and has a diameter of $2.5$ arcsec
($2.3$ kpc), corresponding to the size of the COS aperture.  
In Region B and Bar I, areas with \oiii/\oii$\leq1$ are
visible (coded red in Figure~\ref{fig:ipm}a), indicating that they are
dominated by low-ionization gas.  In particular, Bar I appears to have low
\oiii/\oii, although some islands of higher ionization are clearly
visible.  This is consistent with its reddish appearance in the $UBR$
composite in Figure \ref{fig:rgb}b, suggesting that Bar I
consists of an older stellar population.  In contrast, Bar II, which
is dominated by nebular emission in Figures \ref{fig:rgb}b and
\ref{fig:rgb}c, shows on average high \oiii/\oii\ ratios.  As mentioned
in Section \ref{sec:mu}, this region requires external sources of
ionization, since there are no obvious internal sources that
could provide for the observed high \oiii/\oii\ ratio.  We further
note that two circular regions in the north and south along the major
axis, corresponding to the blue knots in Region B and Bar I, show high
values of \oiii/\oii, but are almost completely surrounded by
low-ionization zones, and are therefore likely optically thick.  Only
the large, circular, ionized central area appears to fully ionize
the ISM to circumgalactic radii, implying that it is the origin of the ionizing photons
responsible for the observed LyC escape.  This region
corresponds to the very blue central Region A of the $UBR$ composite
image in Figure \ref{fig:rgb}, and is approximately indicated by the
white circle in Figure \ref{fig:rgb}a.  As seen in
Figure~\ref{fig:ipm}a, it appears to be even larger when examined via IPM. 

Figures \ref{fig:ipm}b and \ref{fig:ipm}c show high \oiii/\ha\ and
simultaneously low \oii/\ha, spatially coincident with the optically thin
areas and the central galaxy region, confirming our interpretation
of the IPM in Figure \ref{fig:ipm}a.  Figure \ref{fig:ipm}b in
particular shows \oiii\ emission dominant over \ha\ 
throughout most of the galaxy area.  This is also visible in the
three-color image in Figure \ref{fig:rgb}c, which is
completely dominated by emission in the green channel, i.e.,  in
\oiii.  Only the two central knots appear free of both \oiii\ and \ha\
emission in this figure, which has important implications for their
interpretation as optically thin regions (see Section \ref{sec:discussion}). 

As already shown in \citet{Puschnig2017}, the average
  dust extinction in Region A is low. The H$\alpha/$H$\beta$ ratio map in
  Figure \ref{fig:ipm}d shows that the dust extinction outside of Region A is also low. The observed
  high \oiii/\oii\ and low \oii/\ha\ values are therefore not due to high
  internal extinction.

\begin{figure}
  \plotone{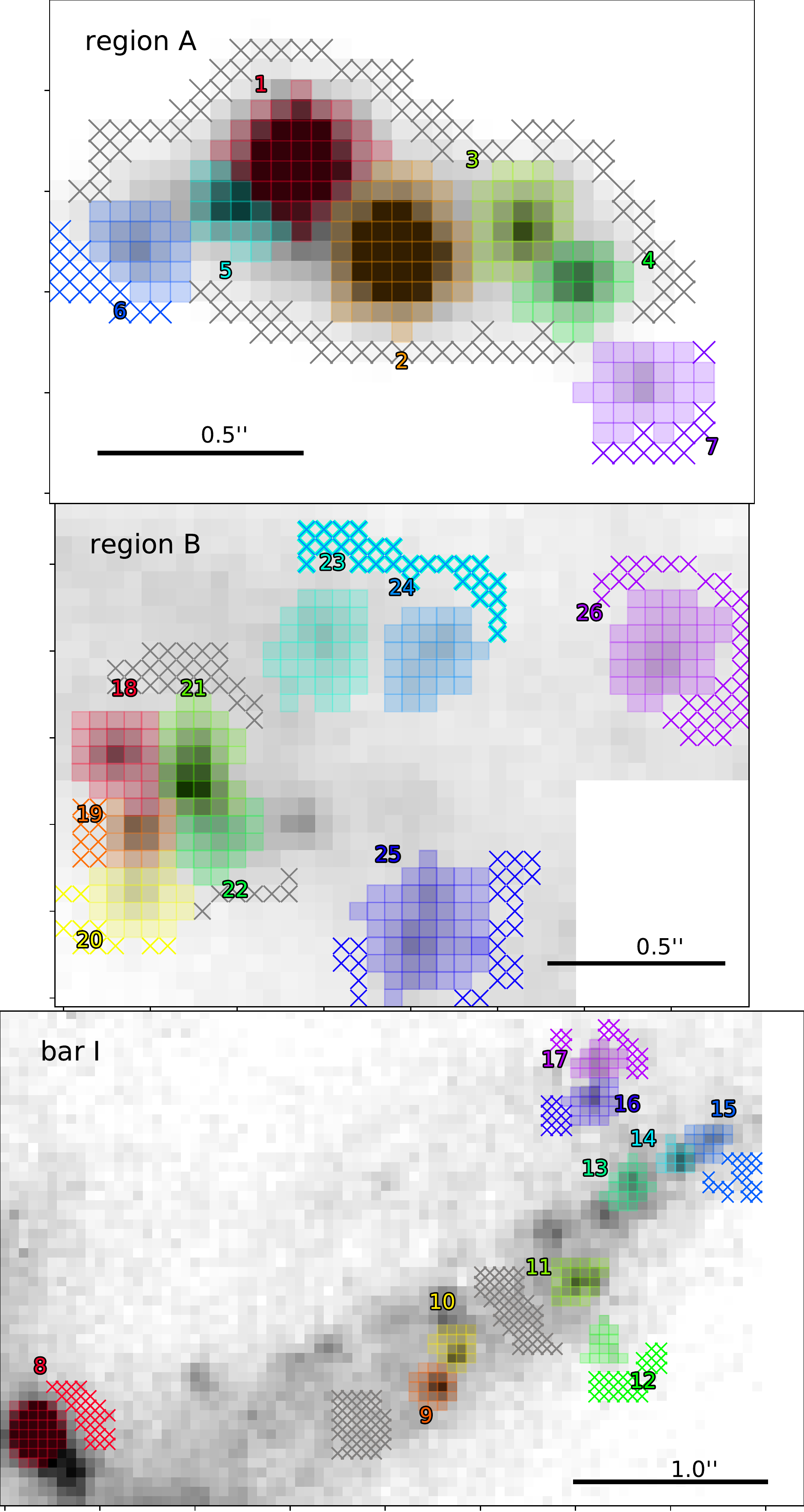}
\caption{FUV continuum (F140LP) images of regions A, B, and
  Bar I, showing cluster IDs.  The colors represent the pixel apertures,
  with corresponding cluster IDs shown in the same color.  
  Colored crosses show sky regions for apertures marked in the same
  color, and gray crosses mark the sky regions common to all clusters
  without an individual sky region.  Clusters 23 and 24 share a sky region.  
  \protect\label{fig:ids}}   
\end{figure}

\section{Stellar Population}\protect\label{sec:phot}
In the context of mapping the galaxy's ionization structure, it is of 
interest to determine the nature of the sources responsible for the
observed ionization.  We therefore examine
the stellar populations that dominate our regions
defined in Figure \ref{fig:rgb}a.
Figure \ref{fig:ids} shows a zoom of the three major areas:  Regions A, B,
and Bar I.  As mentioned earlier, Bar II has no detectable stellar population
and is therefore omitted from this analysis.  Several bright compact knots,
labeled with numbers, are seen in the Figure.  At the luminosity
distance to Tol 1247$-$232 of $213$ Mpc, our
resolution of $0.05$ arcsec corresponds to $47$ pc per pixel.  This is
much larger than the typical sizes for open and globular clusters
($\lesssim10$ pc), and we therefore cannot separate individual star
clusters, unless they happen to be well isolated.  Nevertheless, the
spectral energy distribution (SED) of a knot consisting of several
clusters will likely be dominated by the youngest, brightest population.  For
the remainder of this paper, we therefore refer to these knots as
``clusters'' and treat them as a single stellar population.   

\subsection{Cluster photometry}\protect\label{sec:clusterphot}
For cluster selection criteria, we require at least a $1\sigma$ detection
in all four optical broadband filters, and in the \ha\ filter.  We
discarded objects that showed only noise in the FUV, i.e., older clusters were eliminated since they have negligible contributions to the ionization structure of the gas.
To assess the contribution of the resulting $26$ clusters in Figure \ref{fig:ids} to the
ionization budget of the galaxy, we first obtained their luminosity in
each filter via pixel photometry.  We manually selected pixels
belonging to each cluster (color-coded squares in Figure
\ref{fig:ids}), and then applied a background cutoff to remove pixels
below the selected background level.  The background regions are
marked in Figure \ref{fig:ids} as crosses.  Clusters $1$ through $5$
in Region A all seem to be on the same background, marked by gray
crosses, while clusters $6$ and $7$ in the same region are offset from
the center, and are therefore assigned a different background region each,
marked by crosses of the same color as their pixel aperture.
Similarly, in Region B and Bar I, for all clusters which seemingly
share the same background, the gray crosses mark the location of
the selected background region, while for clusters that are either
isolated or too offset from the main background area, we use separate
regions, marked with crosses color-coded by the corresponding pixel
aperture color.  The median background value of each of these regions
is our estimate for the background flux in each pixel of the
associated cluster apertures. The uncertainties on the photometry
were calculated with the same formula used by the IRAF\footnote{IRAF is
  distributed by the National Optical Astronomy Observatories, 
    which are operated by the Association of Universities for Research
    in Astronomy, Inc., under cooperative agreement with the National
    Science Foundation.} 
 {\tt PHOT} task, using the standard deviation of the sky regions ($stdev$), the flux
 inside the apertures, and the area of both object ($area$) and sky apertures
 ($N_{\rm sky}$), namely
 $\mathrm{err}=\sqrt{ {\rm flux}/{\rm gain} + area\times stdev^2 + area^2\times
   stdev^2/N_{\rm sky}}$. The SED-fitting routine requires symmetric errorbars in the
 input photometry, and we therefore symmetrically propagate the flux error to
 magnitudes (merr) via $\mathrm{merr}=1.0857\times
 \mathrm{err}/\mathrm{flux}$, even though for large uncertainties
 $\gtrsim30\%$ the departure from symmetry is significant. For
   example, a symmetric magnitude error of $0.8$ mag corresponds to a relative
   error of $0.74$, and hence to non-symmetric magnitude errors of
   ${_{-0.60}^{+1.44}}$ mag. Table \ref{tab:phot} summarizes the photometry
 for all $26$ clusters in the available filters. Although in the SED fits we
 use photometry from our own F336W observations and from the repeated F336W
 observations with UVIS2 (PID $13027$), we have omitted explicitly showing the
 latter in the table, because within the error bars it is fully consistent
 with our F336W data. The UVIS2 F336W photometry was included as a separate
 observation in the list provided to the SED-fitting routine.

\begin{deluxetable*}{C|CC|CCCCCCCCCC}
\tabletypesize{\scriptsize}
\tablecaption{Photometry in magnitudes for clusters in Region A, Bar I, and Region B.  \label{tab:phot}}
\tablenum{1}
\tablehead{\colhead{ID} & \colhead{$N_{\rm ap}$} & \colhead{$N_{\rm sky}$} & \colhead{F125LP} & \colhead{F140LP} & \colhead{F336W} & \colhead{FR388N} & \colhead{F438W} & \colhead{FQ508N} & \colhead{FR505N} & \colhead{F547M} & \colhead{F680N} & \colhead{F775W} \\ 
\colhead{} & \colhead{} & \colhead{} & \colhead{(mag)} & \colhead{(mag)} & \colhead{(mag)} & \colhead{(mag)} & \colhead{(mag)} & \colhead{(mag)} & \colhead{(mag)} & \colhead{(mag)} & \colhead{(mag)} & \colhead{(mag)} } 
\startdata
\floattable \\
\multicolumn{13}{c}{Region A}\\
1   &  38  &  82  &  16.9\pm 0.1  &  16.7\pm 0.1  &  17.4\pm 0.1  &  17.7\pm 0.1  &  17.7\pm 0.1  &  17.7\pm 0.1  &  17.1\pm 0.1  &  17.6\pm 0.1  &  17.7\pm 0.1  &  18.0\pm 0.1  \\
2   &  48  &  82  &  17.2\pm 0.1  &  16.9\pm 0.1  &  17.6\pm 0.1  &  17.9\pm 0.1  &  17.7\pm 0.1  &  17.7\pm 0.1  &  17.4\pm 0.1  &  17.5\pm 0.1  &  17.7\pm 0.1  &  17.6\pm 0.1  \\
3   &  25  &  82  &  19.5\pm 0.2  &  19.3\pm 0.2  &  19.9\pm 0.1  &  19.6\pm 0.3  &  20.1\pm 0.1  &  19.6\pm 0.2  &  17.7\pm 0.1  &  19.0\pm 0.1  &  19.0\pm 0.1  &  20.2\pm 0.1  \\
4   &  21  &  82  &  20.2\pm 0.2  &  19.9\pm 0.3  &  20.2\pm 0.1  &  19.9\pm 0.4  &  20.4\pm 0.1  &  19.8\pm 0.2  &  17.7\pm 0.1  &  19.1\pm 0.1  &  19.1\pm 0.1  &  20.6\pm 0.1  \\
5   &  16  &  82  &  19.5\pm 0.2  &  19.2\pm 0.2  &  19.8\pm 0.1  &  19.7\pm 0.3  &  20.0\pm 0.1  &  19.5\pm 0.2  &  17.4\pm 0.1  &  18.8\pm 0.1  &  18.9\pm 0.1  &  20.2\pm 0.1  \\
6   &  21  &  15  &  20.2\pm 0.2  &  20.0\pm 0.3  &  20.8\pm 0.2  &  20.1\pm 0.4  &  20.7\pm 0.2  &  20.1\pm 0.3  &  18.1\pm 0.1  &  19.5\pm 0.1  &  19.4\pm 0.1  &  20.8\pm 0.2  \\
7   &  26  &  16  &  21.0\pm 0.3  &  20.6\pm 0.4  &  21.3\pm 0.2  &  21.2\pm 0.6  &  21.5\pm 0.2  &  21.3\pm 0.5  &  19.7\pm 0.2  &  20.8\pm 0.2  &  20.9\pm 0.3  &  21.7\pm 0.2  \\
\multicolumn{3}{l}{Error [\%] Region A} & 16.6& 19.0& 10.8 &30.0& 10.5& 21.1& 10.2& 7.6& 11.7& 11.5 \\\hline
\multicolumn{13}{c}{Bar I}\\
8   &  37  &  26  &  21.0\pm 0.3  &  20.5\pm 0.3  &  20.3\pm 0.1  &  19.7\pm 0.3  &  20.4\pm 0.1  &  19.7\pm 0.2  &  17.6\pm 0.1  &  19.0\pm 0.1  &  18.9\pm 0.1  &  20.3\pm 0.1  \\
9   &  21  &  81  &  23.0\pm 0.9  &  (22.8)  &  23.1\pm 0.5  &  22.2\pm 1.0  &  23.4\pm 0.5  &  23.0\pm 1.0  &  21.1\pm 0.4  &  22.5\pm 0.4  &  22.2\pm 0.5  &  23.7\pm 0.6  \\
10  &  20  &  81  &  23.4\pm 1.0  &  22.8\pm 1.0  &  23.3\pm 0.5  &  22.2\pm 1.0  &  23.5\pm 0.5  &  22.8\pm 1.0  &  20.9\pm 0.4  &  22.3\pm 0.3  &  22.0\pm 0.4  &  23.5\pm 0.6  \\
11  &  26  &  13  &  23.0\pm 0.8  &  22.5\pm 0.9  &  22.9\pm 0.4  &  (23.0)  &  22.8\pm 0.4  &  22.6\pm 0.9  &  22.2\pm 0.7  &  22.5\pm 0.4  &  22.3\pm 0.5  &  22.4\pm 0.3  \\
12  &  15  &  25  &  (24.4)  &  (23.6)  &  24.5\pm 0.9  &  (23.8)  &  24.4\pm 0.9  &  (23.8)  &  22.7\pm 0.9  &  23.6\pm 0.6  &  23.3\pm 0.8  &  24.0\pm 0.7  \\
13  &  22  &  81  &  (23.9)  &  (23.2)  &  23.2\pm 0.5  &  22.0\pm 0.9  &  23.6\pm 0.6  &  22.8\pm 0.9  &  20.7\pm 0.4  &  22.3\pm 0.3  &  21.8\pm 0.4  &  24.2\pm 0.8  \\
14  &  14  &  23  &  (23.9)  &  (23.2)  &  23.6\pm 0.7  &  (22.8)  &  23.5\pm 0.6  &  22.9\pm 1.0  &  21.0\pm 0.4  &  22.4\pm 0.4  &  22.3\pm 0.5  &  23.4\pm 0.6  \\
15  &  16  &  18  &  (24.0)  &  (23.4)  &  23.6\pm 0.6  &  (23.1)  &  23.7\pm 0.6  &  (23.4)  &  22.2\pm 0.7  &  23.1\pm 0.5  &  22.8\pm 0.7  &  23.3\pm 0.5  \\
16  &  23  &  11  &  (24.0)  &  (24.0)  &  23.7\pm 0.7  &  (23.4)  &  24.0\pm 0.7  &  (23.7)  &  22.5\pm 0.8  &  23.3\pm 0.5  &  23.4\pm 0.9  &  24.0\pm 0.7  \\
17  &  23  &  19  &  (23.9)  &  (23.4)  &  23.9\pm 0.7  &  (23.3)  &  24.0\pm 0.7  &  (23.9)  &  22.7\pm 0.9  &  23.6\pm 0.6  &  23.3\pm 0.8  &  23.7\pm 0.6  \\
\multicolumn{3}{l}{Error [\%] Bar I}    & 61.1& 56.3& 49.1& 57.2& 51.6& 68.8& 51.8& 36.9& 51.9& 50.5 \\\hline
\multicolumn{13}{c}{Region B}\\
18  &  27  &  34  &  22.6\pm 0.7  &  22.2\pm 0.7  &  22.3\pm 0.3  &  21.5\pm 0.7  &  22.3\pm 0.3  &  21.7\pm 0.6  &  19.4\pm 0.2  &  20.9\pm 0.2  &  20.8\pm 0.3  &  22.4\pm 0.3  \\
19  &  13  &  8   &  (23.5)  &  (22.8)  &  22.7\pm 0.4  &  (22.3)  &  22.6\pm 0.4  &  21.8\pm 0.6  &  19.7\pm 0.2  &  21.0\pm 0.2  &  20.9\pm 0.3  &  22.2\pm 0.3  \\
20  &  22  &  12  &  (23.7)  &  (23.1)  &  23.3\pm 0.6  &  22.2\pm 1.0  &  23.2\pm 0.5  &  22.3\pm 0.8  &  19.9\pm 0.2  &  21.5\pm 0.2  &  21.3\pm 0.3  &  22.8\pm 0.4  \\
21  &  29  &  34  &  22.1\pm 0.5  &  21.6\pm 0.6  &  21.6\pm 0.2  &  21.4\pm 0.7  &  21.6\pm 0.2  &  21.5\pm 0.5  &  20.0\pm 0.3  &  21.0\pm 0.2  &  21.0\pm 0.3  &  21.5\pm 0.2  \\
22  &  19  &  34  &  (23.4)  &  (22.8)  &  22.7\pm 0.4  &  21.9\pm 0.9  &  22.8\pm 0.4  &  22.1\pm 0.7  &  20.0\pm 0.3  &  21.4\pm 0.2  &  21.2\pm 0.3  &  22.5\pm 0.4  \\
23  &  35  &  30  &  23.0\pm 0.8  &  22.3\pm 0.8  &  22.8\pm 0.4  &  21.6\pm 0.8  &  22.6\pm 0.3  &  21.9\pm 0.6  &  20.4\pm 0.3  &  21.5\pm 0.2  &  21.3\pm 0.3  &  22.2\pm 0.3  \\
24  &  34  &  30  &  23.1\pm 0.8  &  22.5\pm 0.8  &  22.9\pm 0.4  &  21.8\pm 0.8  &  23.1\pm 0.4  &  22.2\pm 0.7  &  20.3\pm 0.3  &  21.7\pm 0.2  &  21.5\pm 0.3  &  22.9\pm 0.4  \\
25  &  53  &  30  &  22.9\pm 0.8  &  22.3\pm 0.8  &  22.5\pm 0.4  &  (22.2)  &  22.3\pm 0.3  &  21.8\pm 0.6  &  19.8\pm 0.2  &  21.3\pm 0.2  &  21.2\pm 0.3  &  22.2\pm 0.3  \\
26  &  40  &  32  &  23.0\pm 0.8  &  22.7\pm 0.9  &  23.2\pm 0.5  &  (22.1)  &  23.1\pm 0.4  &  22.4\pm 0.8  &  20.4\pm 0.3  &  21.8\pm 0.3  &  21.7\pm 0.4  &  23.2\pm 0.4  \\
\multicolumn{3}{l}{Error [\%] Region B} & 67.9& 70.6& 38.1& 75.4& 34.3& 60.3& 23.1& 19.5& 28.6& 31.5 \\
\enddata
\tablecomments{Columns 2 and 3 give the number of pixels in each
  aperture ($N_{\rm ap}$) and in the associated sky region
  ($N_{\mathrm{sky}}$). \\
  Non-detections, i.e. observations with fractional 
  error $\ge100\%$, are treated as upper limits and the $1\sigma$ value is
  instead given in parentheses. \\
  Symmetric magnitude errors are
  shown. Symmetric errors of $0.6, 0.8, 1.0$ mag correspond to actual,
  non-symmetric errors of ${_{-0.5}^{+0.9}, {}_{-0.6}^{+1.4},
    {}_{-0.7}^{+2.8}} $ mag,\\
  respectively. Galactic reddening correction has
  been applied. The bottom row of each region shows the average percentage
  uncertainty \\ 
  (relative error in per cent) for each filter.} 
\end{deluxetable*}
\normalsize

Aperture corrections for each filter and each cluster were obtained
with {\tt pysynphot}, by assuming that the correction for the
corresponding pixel aperture is equivalent to that of a circular
aperture with radius $r=\sqrt{\textrm{area}/\pi}$. These corrections account
for variations in the point spread function (PSF) in the different filters,
although due to the non-circular geometry of the pixel apertures, these
corrections may be somewhat overestimated. Further, all photometry has been
corrected for Galactic extinction. The \citet{Schlafly2011} reddening curve
was used for all $HST$ filters redwards of F336W.  For the F125LP and F140LP
filters we instead used the \citet{Sasseen2002} FUV attenuation curve. 

\begin{deluxetable}{CCCCCCC}
\tabletypesize{\scriptsize}
\tablecaption{Cluster SED parameters from Cigale fits. \label{tab:cigale}}
\tablenum{2}
\tablehead{\colhead{ID} & \colhead{Age} & \colhead{$f_{\rm esc}$} & \colhead{$E(B-V)$} & \colhead{$Q\mathrm{(H^0)}$} & \colhead{$\mathrm{M_\star}$} & \colhead{$\chi^2_\nu$} \\ 
\colhead{} & \colhead{($\mathrm{Myr}$)} & \colhead{} & \colhead{(mag)} & \colhead{($\mathrm{10^{52}\,s^{-1}}$)} & \colhead{($\mathrm{10^7\,M_{\odot}}$)} & \colhead{} } 
\startdata
\multicolumn{7}{c}{Region A}\\
     1 &  4_{-0}^{+2}  &    0.95_{- 0.10}^{+ 0.00}   &   0.05_{- 0.05}^{+ 0.01} &      151.02_{-114.03}^{+9.39} &        6.56_{-0.82}^{+0.41} &  0.64 \\
     2 &  12_{-1}^{+7} &    0.95_{- 0.95}^{+ 0.00}   &   0.00_{- 0.00}^{+ 0.02} &        3.45_{-2.77}^{+1.00}   &       12.16_{-1.86}^{+7.89} &  0.98 \\
     3 &  2_{-0}^{+2}  &    0.80_{- 0.30}^{+ 0.05}   &   0.18_{- 0.09}^{+ 0.04} &       76.85_{-59.71}^{+24.60} &        2.77_{-2.02}^{+0.89} &  0.32 \\
     4 &  2_{-0}^{+1}  &    0.75_{- 0.10}^{+ 0.00}   &   0.21_{- 0.03}^{+ 0.01} &       64.71_{-16.44}^{+3.75}  &        2.33_{-0.59}^{+0.14} &  0.18 \\
     5 &  2_{-0}^{+1}  &    0.80_{- 0.00}^{+ 0.00}   &   0.16_{- 0.01}^{+ 0.02} &       76.60_{-4.13}^{+8.51}   &        2.76_{-0.15}^{+0.31} &  0.24 \\
     6 &  4_{-0}^{+0}  &    0.35_{- 0.20}^{+ 0.10}   &   0.10_{- 0.02}^{+ 0.02} &        8.99_{-1.89}^{+1.81}   &        0.39_{-0.08}^{+0.08} &  0.16 \\
     7 &  2_{-0}^{+1}  &    0.90_{- 0.05}^{+ 0.00}   &   0.18_{- 0.03}^{+ 0.02} &       23.44_{-4.97}^{+2.87}   &        0.84_{-0.18}^{+0.10} &  0.10 \\
\multicolumn{7}{c}{Bar I}\\
     8 &  2_{-0}^{+1}  &    0.75_{- 0.00}^{+ 0.00}   &   0.32_{- 0.00}^{+ 0.00} &      105.82_{-0.00}^{+0.00}   &        3.81_{-0.00}^{+0.00} &  0.01 \\
     9 &  2_{-0}^{+1}  &    0.90_{- 0.00}^{+ 0.00}   &   0.20_{- 0.01}^{+ 0.01} &        4.79_{-0.27}^{+0.29}   &        0.17_{-0.01}^{+0.01} &  0.02 \\
    10 &  2_{-0}^{+1}  &    0.80_{- 0.10}^{+ 0.00}   &   0.23_{- 0.03}^{+ 0.03} &        4.25_{-1.03}^{+0.82}   &        0.15_{-0.04}^{+0.03} &  0.05 \\
    11 &  12_{-1}^{+0} &    0.95_{- 0.95}^{+ 0.00}   &   0.14_{- 0.04}^{+ 0.01} &        0.07_{-0.01}^{+0.02}   &        0.23_{-0.07}^{+0.01} &  0.04 \\
    12 &  11_{-1}^{+1} &    0.00_{- 0.00}^{+ 0.90}   &   0.09_{- 0.03}^{+ 0.04} &        0.02_{-0.00}^{+0.01}   &        0.04_{-0.01}^{+0.02} &  0.04 \\
    13 &  2_{-0}^{+1}  &    0.80_{- 0.05}^{+ 0.10}   &   0.22_{- 0.04}^{+ 0.05} &        3.97_{-1.02}^{+1.67}   &        0.14_{-0.04}^{+0.06} &  0.25 \\
    14 &  2_{-0}^{+1}  &    0.85_{- 0.00}^{+ 0.00}   &   0.35_{- 0.01}^{+ 0.02} &        7.51_{-0.43}^{+0.93}   &        0.27_{-0.02}^{+0.03} &  0.01 \\
    15 &  4_{-2}^{+14} &    0.95_{- 0.95}^{+ 0.00}   &   0.27_{- 0.13}^{+ 0.11} &        2.08_{-2.08}^{+7.45}   &        0.09_{-0.02}^{+0.25} &  0.06 \\
    16 &  2_{-0}^{+1}  &    0.95_{- 0.00}^{+ 0.00}   &   0.29_{- 0.02}^{+ 0.01} &        4.81_{-0.54}^{+0.30}   &        0.17_{-0.02}^{+0.01} &  0.03 \\
    17 &  12_{-0}^{+7} &    0.15_{- 0.15}^{+ 0.80}   &   0.08_{- 0.02}^{+ 0.02} &        0.02_{-0.01}^{+0.00}   &        0.06_{-0.01}^{+0.05} &  0.03 \\
\multicolumn{7}{c}{Region B}\\                                         
    18 &  2_{-0}^{+1}  &    0.70_{- 0.05}^{+ 0.00}   &   0.27_{- 0.02}^{+ 0.02} &       13.72_{-2.19}^{+1.60}   &        0.49_{-0.08}^{+0.06} &  0.05 \\
    19 &  2_{-0}^{+1}  &    0.70_{- 0.05}^{+ 0.05}   &   0.38_{- 0.02}^{+ 0.04} &       19.44_{-3.03}^{+6.40}   &        0.70_{-0.11}^{+0.23} &  0.09 \\
    20 &  2_{-0}^{+2}  &    0.50_{- 0.50}^{+ 0.15}   &   0.34_{- 0.09}^{+ 0.07} &        6.99_{-5.33}^{+5.09}   &        0.25_{-0.18}^{+0.18} &  0.08 \\
    21 &  2_{-0}^{+2}  &    0.95_{- 0.05}^{+ 0.00}   &   0.34_{- 0.09}^{+ 0.00} &       46.63_{-35.93}^{+0.15}  &        1.68_{-1.21}^{+0.01} &  0.02 \\
    22 &  2_{-0}^{+1}  &    0.80_{- 0.05}^{+ 0.00}   &   0.35_{- 0.03}^{+ 0.02} &       15.87_{-3.26}^{+1.95}   &        0.57_{-0.12}^{+0.07} &  0.06 \\
    23 &  4_{-0}^{+0}  &    0.65_{- 0.40}^{+ 0.05}   &   0.33_{- 0.06}^{+ 0.03} &        5.85_{-2.47}^{+1.26}   &        0.25_{-0.11}^{+0.05} &  0.13 \\
    24 &  2_{-0}^{+1}  &    0.70_{- 0.20}^{+ 0.05}   &   0.25_{- 0.05}^{+ 0.05} &        6.31_{-2.43}^{+2.55}   &        0.23_{-0.09}^{+0.09} &  0.09 \\
    25 &  2_{-0}^{+1}  &    0.85_{- 0.05}^{+ 0.00}   &   0.36_{- 0.04}^{+ 0.02} &       23.67_{-6.07}^{+2.68}   &        0.85_{-0.22}^{+0.10} &  0.07 \\
    26 &  4_{-0}^{+0}  &    0.35_{- 0.10}^{+ 0.05}   &   0.16_{- 0.02}^{+ 0.01} &        1.34_{-0.18}^{+0.13}   &        0.06_{-0.01}^{+0.01} &  0.03 \\
\enddata
\tablecomments{Uncertainties correspond to the resulting parameter range for\\
 models within $20\%$ of the minimum $\chi^2$.}
\end{deluxetable}
\normalsize

We carry out pixel photometry because the regions are too
crowded to allow for aperture photometry with a fixed radius.
Using small
apertures to accomodate the small distances between clusters resulted
in large, and therefore uncertain, aperture corrections on the order
of  $\sim1$ mag.  We also attempted to model each cluster with a 2D Moffat
function with \textsc{astropy} \citep{AstropyCollaboration2013}.  The
model photometry of the brightest clusters was consistent with the
pixel photometry within the uncertainties, while the fainter clusters
were difficult to model.  Our pixel photometry for the clusters is
summarized in Table \ref{tab:phot}.  

\subsection{Cluster SED modeling}\protect\label{sec:seds}
Once the photometry was obtained, we modeled the SED of each cluster
with the Cigale software \citep[v 0.11.0, ][]{Noll2009, Serra2011}.
The available $HST$ filters provide $11$ photometry measurements
covering the FUV and optical from the $10$ filters listed in Section
\ref{sec:data}, and one repeated observation in F336W from a separate
observing program. Cigale accounts for nebular emission by adding a
  generic \hii\ region spectrum, which may or may not be representative of
  the physical conditions in Tol 1247-232. Further, the nebular emission in
  each cluster aperture may be due to ionizing sources outside of the
  aperture. For these reasons, we treat the narrowband photometry as upper
  limits during the SED fitting.
In addition, to account for the possibility that the narrowband photometry
underestimates the flux in the emission lines due to escape of
ionizing photons beyond the aperture, the $f_\mathrm{esc}$ parameter was allowed to vary
between $0.0$ and $1.0$ in $0.05$ steps for each cluster. We chose a
Salpeter \citep{Salpeter1955} initial mass function (IMF) with
$0.6\leq\mathrm{M}\leq120 \rm M_\odot$, and a constant metallicity of
$Z=0.004$.  This metallicity is consistent with that measured for Tol
1247$-$232 by \citet{Terlevich1993}, who used the so-called ``direct''
method based on the determination of the electron temperature via
detection of the temperature-sensitive [O {\small III}] line at
$\lambda 4363$ \AA.  We further assumed a quasi-instantanous star
formation history (SFH) in the form of a delayed decreasing exponential with
time $t\times e^{-t/\tau_{\rm SFH}}$, adopting a very short 
characteristic timescale $\tau_{\rm SFH}=0.01$ Myr. During the fit, the age of the
stellar population was allowed to vary from $2$ to $100$ Myr, in steps
of $1$ Myr, which is the smallest step size allowed by Cigale. The
ionization parameter was allowed to take on values of $-4 \leq \log U \leq -1$ in steps of $-0.1$.  The \citet{Calzetti2000} dust attenuation
law was used to fit the internal $E(B-V)$, which was
allowed to vary from $0.0$ to $0.7$ in steps of $0.01$. Due to lack of IR data
with sufficient resolution, we cannot constrain the dust emission for the
clusters, and therefore the assumed fraction of ionizing photons absorbed by dust, $f_\mathrm{dust}$, is set to
$0.0$ and kept constant during the fit. Therefore the individual cluster escape
fractions $f_{\rm esc}$ in Table \ref{tab:cigale} are an upper limit, in the absence of any dust inside
of the \hii\ regions. 

\normalsize
\begin{figure*}
  \plotone{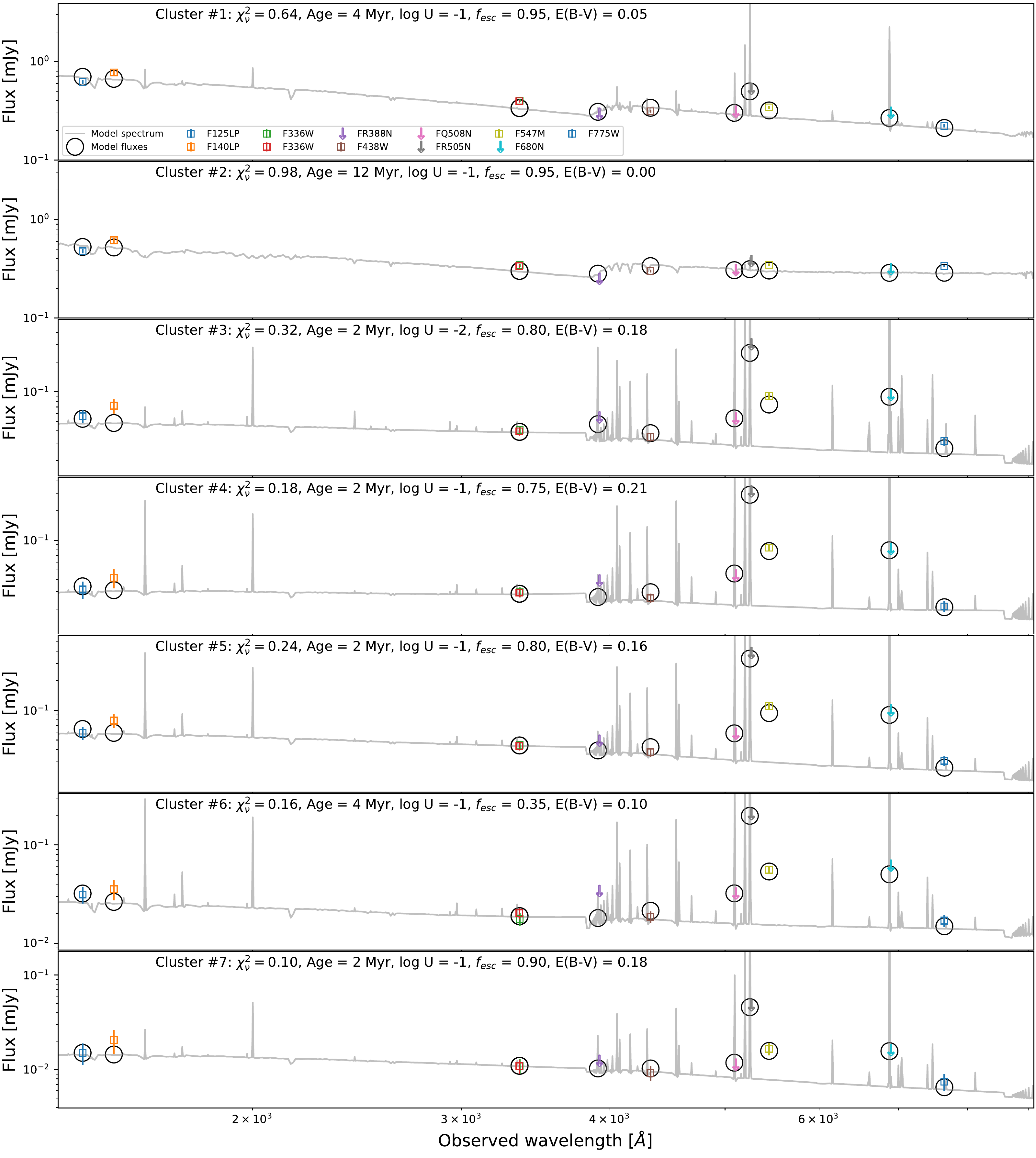}
\caption{Cigale SED fits for Region A.  Photometric data are shown with
  circles, upper limits are indicated by arrows.  \protect\label{fig:sed1}}  
\end{figure*}

To break the age-extinction degeneracy, \citet{Fouesneau2012}
recommend using an \ha\ filter in addition to broadband $UBVI$
filters.  Our cluster detection criteria therefore required a
$\geq1\sigma$ detection in $UBVR$ and the \ha\ filter.  Table
\ref{tab:phot} lists the average percentage uncertainty for all
regions and all filters.  Region A is the brightest, strongly
dominated by Clusters 1 and 2, followed by Region B and Bar I.  The latter contains
some of the faintest clusters in the sample.  We are predominantly
interested in Region A, since it dominates the ionization budget of
the galaxy, as we will show in Section \ref{sec:discussion}.  The
uncertainties in all $10$ filters are here on average $\leq30$\%,
and hence the SED is well constrained.
For Region B and Bar I, the average uncertainties are $\lesssim38$\%
and $52$\%, respectively, for the optical broadband and the \ha\
filters.  In the FUV, Region B and Bar I are not as well constrained, with
$\lesssim71$\% uncertainty and several non-detections among the clusters. 
 The situation is similar for both of these regions for
the narrowband filters FR388N (\oii) and FQ508N (\hb).  The large
uncertainties in these filters are of little consequence, since we use
all narrowband photometry as upper limits in the SED fits, as
described above.  For Regions 
A and B we can test the robustness of our results by also performing
SED fits using only the broadband data, which include two FUV filters,
and thus one can still break the age-extinction degeneracy without the
\ha\ filter.  Within the uncertainties, the results were consistent
with the SED fits using all available filters (FUV, optical broad- and
narrowbands), and we therefore proceed with the SED fits using all
filters. 

Table \ref{tab:cigale} shows the major parameters from the best fit
models, namely age, $E(B-V)$, $f_\mathrm{esc}$, production rate of
ionizing photons $Q\mathrm{(H^0)}$, and stellar mass
${M_\star}$. In Figure~\ref{fig:sed1} we show the corresponding best SED fits for Region A,
while Figures \ref{fig:sedbarI} and \ref{fig:sedB} of the Appendix
show the SED fits for the remaining clusters in Bar I and Region B,
respectively.  The upper (lower) uncertainties on the presented parameters are
simply the difference between the parameter value at minimum $\chi^2$ and the
maximum (minimum) value of the parameter range obtained from all models with
$\chi^2$ within $20\%$ of the minimum. This is illustrated in Figures
\ref{fig:pdf_age} and \ref{fig:pdf_fesc} of the Appendix for the age and
$f_\mathrm{esc}$ parameters, respectively.

\begin{figure}
  \gridline{
    \fig{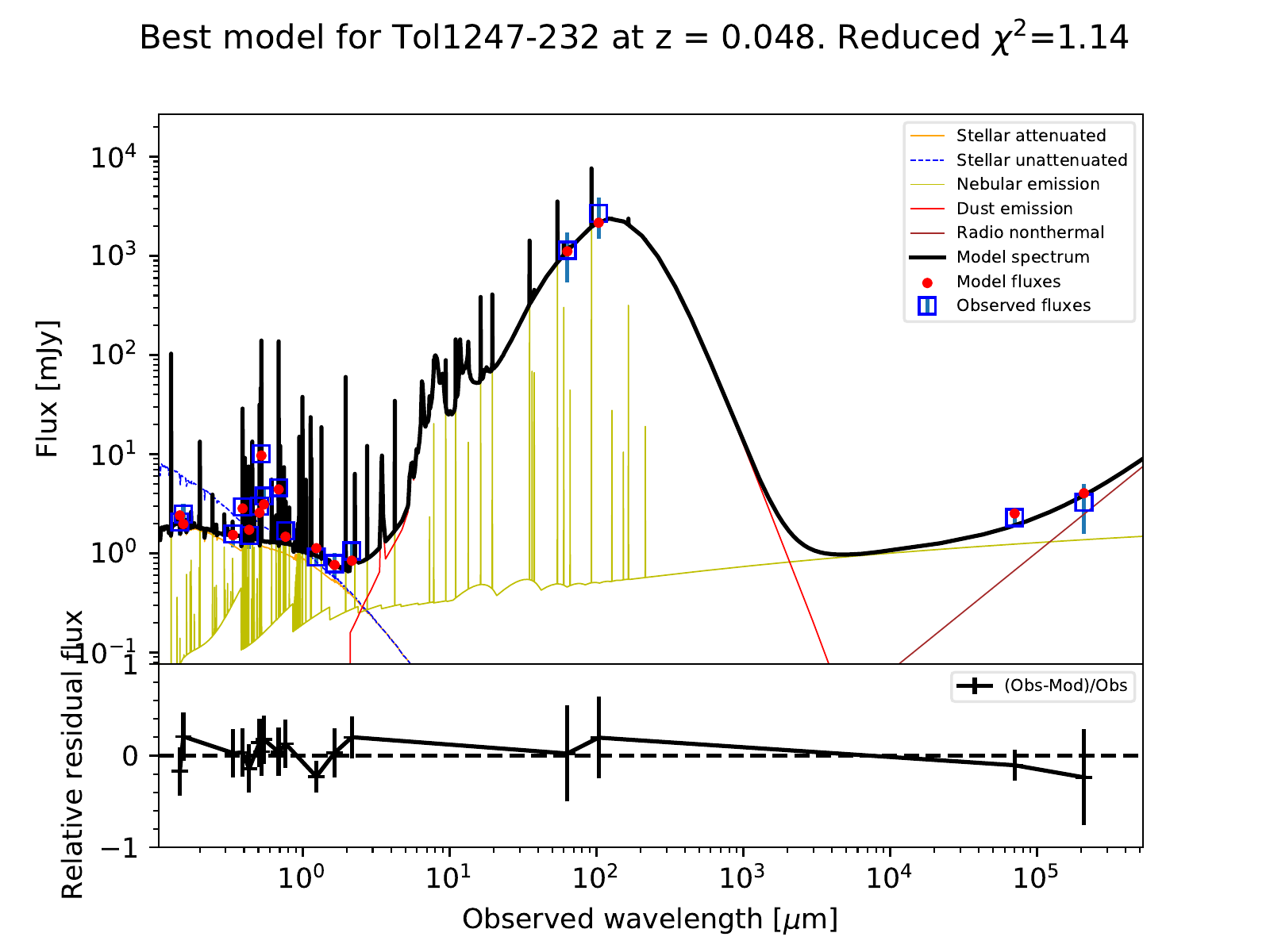}{0.5\textwidth}{(a)}
  }
  \gridline{
    \fig{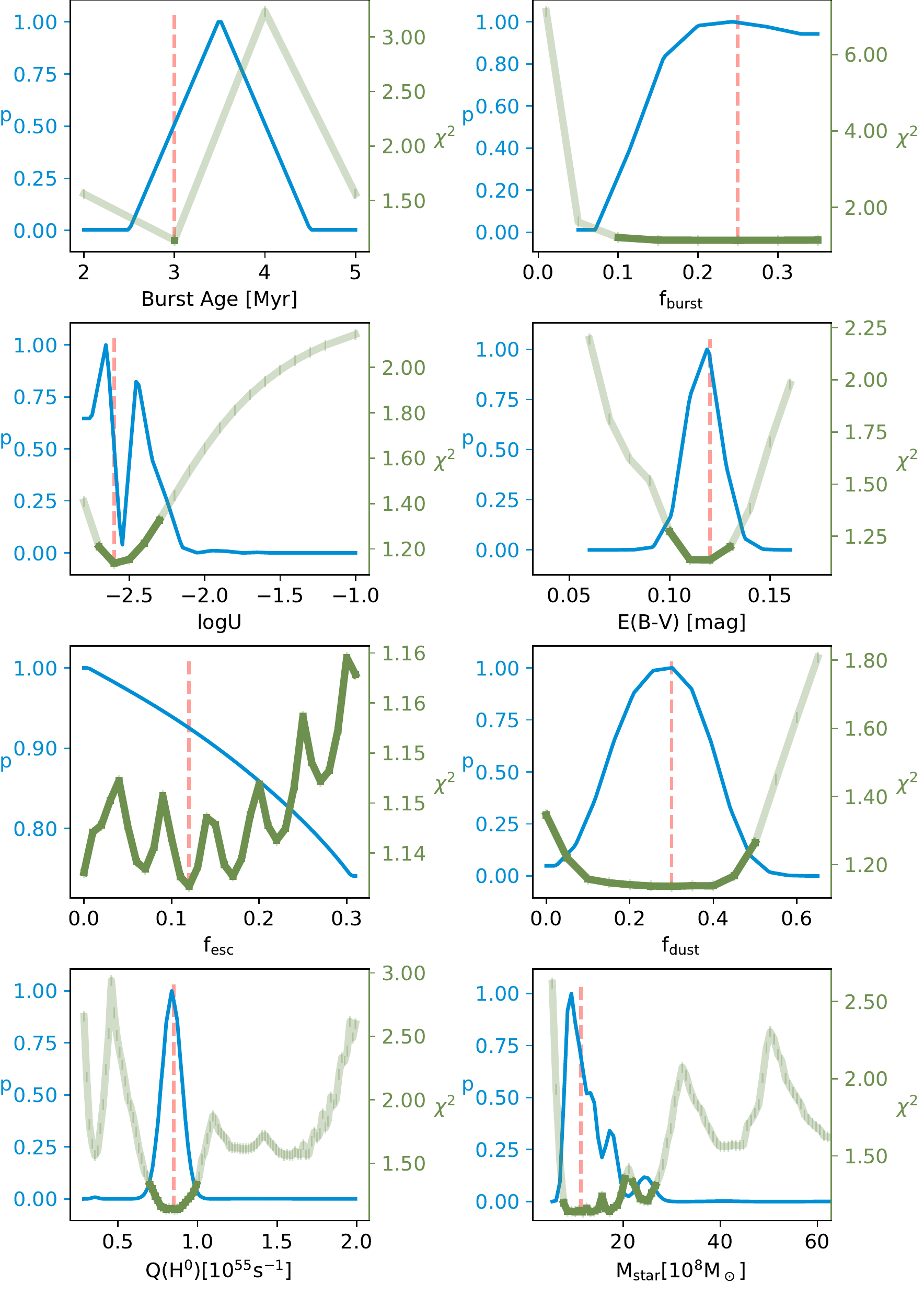}{0.5\textwidth}{(b)}
  }
\caption{(a) Log plot of Cigale SED fit of Tol 1247$-$232, with fit components
  as shown in the legend.  (b) PDFs of the major fitted parameters, burst age,
  young mass fraction, $f_{\rm burst}$, $\log U$, 
  $E(B-V)$, $f_{\rm esc}$, $f_{\rm dust}$, $Q$(H$^0$), and $M_\star$. The
  PDFs are color-coded in  blue, labeled on the left $y$-axes; the
  $\chi^2$ distributions are color-coded in green, labeled on
  the right $y$-axes. The best model values selected by Cigale, corresponding
  to the lowest $\chi^2$, are marked by dashed red lines.  $\chi^2$ values
  within $20\%$ of this minimum represent models indistinguishable from the
  best fit, and are marked by dark green.\protect\label{fig:tol1247}} 
\end{figure}

We note several things about the SED fits and their uncertainties.
First, in some cases the maximum probability does not correspond to
the minimum $\chi^2$ as seen in Figures \ref{fig:pdf_age} and
\ref{fig:pdf_fesc}.  This is expected, because in each parameter bin,
the probability is a weighted sum, evaluated over the $\chi^2$ values
of all models in that bin.  Therefore, while the best model is in the
minimum $\chi^2$ bin, other bins may contain several good models,
which may increase their probability enough to offset the peak of the
probability density function (PDF) \citep{Noll2009}.  Second, Cigale computes
the reduced chi-square statistic, $\chi^2_\nu$, where $\nu$ is the number
of degrees of freedom, and $\chi^2_\nu=\chi^2/\nu$. In principle,
$\chi^2_\nu>1$ indicates
either that the error variance of the data has been underestimated or
that the model is not fully capturing the data, while $\chi^2_\nu<1$
indicates either that the model is fitting noise or that the error
variance has been overestimated \citep[e.g.][]{Bevington1969}.
The latter is likely the reason for the models with $\chi^2_\nu<1$ in
Table \ref{tab:cigale} and Figures \ref{fig:pdf_age} and
\ref{fig:pdf_fesc}, since Cigale cannot be fitting noise in our setup.
Third, even though Cigale selects a best value for a given parameter,
that value may be poorly constrained.  Examination of the PDF and
$\chi^2$ distribution is invaluable in identifying such cases.
For four clusters in Table \ref{tab:cigale} (Clusters 2, 11, 12, 17) the
escape fraction parameter is unconstrained, with possible values covering
$0-100$\% escape.  To illustrate the stablility of the model parameters, we
show the PDF and $\chi^2_\nu$ distributions for the age and the escape
fraction in Figures \ref{fig:pdf_age} and \ref{fig:pdf_fesc}. Lastly, Cigale
is designed primarily for stellar population synthesis modeling of integrated
galaxy populations, rather than individual clusters.  Since the distribution
of nebular light associated with individual clusters corresponds poorly to the
spatial apertures of the clusters, nebular parameters such as log U fitted by
Cigale are not meaningful, and therefore not presented in Table
\ref{tab:cigale}.

\subsection{Galaxy SED modeling}\protect\label{sec:seds_galaxy}
For the $26$ clusters we have used only FUV and optical data, because
these data have the spatial resolution to separate individual ionizing
sources.  For the SED of the entire galaxy we can use integrated
values from observations with lower resolution, namely, the photometry
of the integrated galaxy area in the near-infrared (NIR) \Jband,
\Hband, and \Kband\ bands from the 2MASS catalog, in the infrared (IR)
at $60$ and $100\mu$m from the IRAS catalog, and at $1.49$ GHz and
$4.8$ GHz in the radio from \citet{RosaGonzalez2007}. To obtain the
integrated FUV and optical photometry of the entire galaxy we
performed aperture photometry on all FUV and optical images, with a
radius of $5.8$ arcsec, as indicated in Figure \ref{fig:rgb}a. Since
    we are now evaluating the entire stellar population of Tol 1247$-$232, we
    assumed a double exponential SFH law, with one exponential for the young
    population, and one for the underlying old population. We assumed an old
    population of $6$ Gyr, with a short $e$-folding time
    $\mathrm{\tau_{old}}=0.01$ Myr. For the superimposed young population, we
    assumed an equally short $e$-folding time $\mathrm{\tau_{burst}}=0.01$
    Myr, and varied the young mass fraction
    $f_{\rm burst}$ between $0.01$ and $0.35$ in steps of $0.05$, and the
    burst age between $2$ to $20$ Myr in steps of $1$ Myr. In addition to
    these, the $\log U$, $f_{\rm esc}$, $f_{\rm dust}$ and $E(B-V)$
    parameters were also varied with the same range as for the clusters.
    The best model parameters
    are listed in Table \ref{tab:cigale2}, with the SED displayed in Figure
    \ref{fig:tol1247}a. The PDF of the major parameters and the minimum
    $\chi^2_\nu$ are shown in Figure \ref{fig:tol1247}b. 

\begin{deluxetable*}{CCCCCCCCCC}
\tabletypesize{\scriptsize}
\tablecaption{Tol 1247-232 SED parameters from Cigale fits, using a double   exponential SFH. \label{tab:cigale2}}
\tablenum{3}
\tablehead{\colhead{ID} & \colhead{Burst Age} & \colhead{$f_{\rm burst}$} & \colhead{log $U$} & \colhead{$f_{\rm esc}$} & \colhead{$f_{\rm dust}$} & \colhead{$E(B-V)$} & \colhead{$Q\mathrm{(H^0)}$} & \colhead{$\mathrm{M_\star}$} & \colhead{$\chi^2_\nu$} \\ 
\colhead{} & \colhead{($\mathrm{Myr}$)} & \colhead{} & \colhead{} & \colhead{} & \colhead{} & \colhead{} & \colhead{($\mathrm{10^{54}\,s^{-1}}$)} & \colhead{($\mathrm{10^9\,M_{\odot}}$)} & \colhead{} } 
\startdata
\mathrm{Tol}\ 1247$-$232  &  3_{-0}^{+0}  &  0.25_{-0.15}^{+0.10}  &  -2.6_{-0.1}^{+0.3}  &  0.12_{-0.12}^{+0.19} & 0.3_{-0.3}^{+0.2}  & 0.12_{-0.02}^{+0.01}  &  8.49_{-1.60}^{+1.36}     & 1.13_{-0.40}^{+1.53} & 1.14 \\
\enddata
\tablecomments{Uncertainties correspond to the resulting parameter range for models within $5\%$ of the minimum $\chi^2$.}
\end{deluxetable*}
\normalsize

The IR data constrain the dust content and we can therefore fit the
fraction of ionizing photons absorbed by dust, $f_\mathrm{dust}$.  For
the best model, and models with similar $\chi^2_\nu$,
$f_\mathrm{dust}=0.3_{-0.3}^{+0.2}$.  Typical fractions of
dust-absorbed LyC photons are $\sim50\%$ for solar and LMC
metallicities \citep{Inoue2001}.  The lower, SMC-like, metallicity of
Tol 1247$-$232 is consistent with $f_\mathrm{dust}$ fractions being somewhat
lower here, although within the uncertainties, $f_\mathrm{dust}$ is also consistent
with fractions for solar and LMC metallicities.  We obtain a dust attenuation
of $E(B-V)=0.12_{-0.02}^{+0.01}$, which is consistent with estimates from
observed Balmer line ratios \citep[$E(B-V)=0.13$;][]{Puschnig2017} and
with modeling the SED by fitting the observed FUV COS spectrum
\citep[$E(B-V)=0.11$;][]{Leitherer2016}. 

Within the uncertainties, the resulting stellar mass
  $1.13_{-0.40}^{+1.53}\times10^9\mathrm{M_\odot}$ is a factor of $2.2$ from
  the estimate by \citet[$5.9\times10^9\mathrm{M_\odot}$;][]{Leitet2013}, who
obtained the stellar mass by simply assuming a reasonable mass-to-light ratio
based on the statistical average of local SFGs, and hence their
  estimate can  certainly be off by a factor of $2\mbox{-}3$. Comparing our
mass to the total mass of all clusters in Table \ref{tab:cigale}, we find that
the young mass fraction is $\sim34\%$, which is consistent within the
errorbars with the value found by Cigale. This young population gives a total
production  rate of ionizing photons $Q\mathrm{(H^0)}_{\rm
  total}=8.5\times10^{54}$ s$^{-1}$. From these, a fraction of $0.42$ are
absorbed by dust and/or escape (Table \ref{tab:cigale2}), and the remaining
$Q\mathrm{(H^0)}_{\rm ion}=4.9\times10^{54}$ photons per second are left to
ionize the gas. Note, that the escape fraction is unconstrained, in the sense
that all values below $0.31$ are consistent with the best model
(Table~\ref{tab:cigale2}).  

\subsection{Classical WR or VMS Stars?}\protect\label{wr_vms}

Wolf-Rayet (WR) stars in Tol 1247$-$232 have been reported by
\citet{Masegosa1991} and confirmed by \citet{Schaerer1999} through re-analysis
of the same data.  Classical WR stars of the carbon (WC) and nitrogen
(WN) sequences are the stripped cores of evolved massive stars.  Their strong stellar winds
provide substantial mechanical feedback, and they can be used to age-date stellar
clusters, implying an age of $\sim5$ Myr \citep[e.g.,][]{Crowther2007}.  WN
stars can be inferred from the so-called ``blue bump'' due to broad He
{\small II} $\lambda4686$ emission, while WC stars will display a ``red bump''
at $\lambda5810$ \AA.   
 
An alternative explanation to the ``blue bump'' may be very massive stars
(VMS).  Young VMS are known to also show WN spectral
features \citep[e.g.][]{Crowther2010,Crowther2011,Grafener2015,Smith2016}.  VMS
O-type stars have been detected in the LMC SSC R136, a very massive
and extremely young cluster \citep[$\mathrm{M\geq10^4}\ M_\odot$, $\leq2$
  Myr;][]{Crowther2010}.  A similar situation has been suggested in Mrk 71
\citep{James2016, Micheva2017}, where VMS could explain the detection of broad
He {\small II} and are consistent with the $\sim1$ Myr age of its dominant SSC,
Knot A.  The SED models in our analysis cannot differentiate ages $\leq3$
Myr, and some of the clusters in Tol 1247$-$232 could be even younger than what
we have indicated in Table \ref{tab:cigale}.  It is therefore possible that
the observed WN spectral features are due to VMS stars in SSCs
of extremely young ages $\sim1$ Myr in this galaxy.  

The presence of WC stars would support the interpretation of a
classical WN population.  We evaluated the possibility of WC stars by
re-examining  the available spectra.  There are two optical spectra of Tol 1247$-$232  
from the ESO $3.6$m and the Las Campanas DuPont telescopes, published in the  
\hii\ galaxy catalog of \citet{Terlevich1991}, and used to detect the blue  
bump in \citet{Masegosa1991} and \citet{Schaerer1999}.   These data were  
kindly provided to us by R.  J.  Terlevich.  Due to  
the low resolution of the ESO spectrum, and a second order contamination of  
the DuPont spectrum (R.  J.  Terlevich, private communication), we are unable to  
definitively exclude the presence of the $\lambda5810$\AA\ red bump
and implied WC stars.  If a red bump is present in these spectra, it
is below the detection limit. Therefore, the WR stars in Tol 1247$-$232 are likely
dominated by WN type, and the possibility of VMS cannot be discarded.

Cigale assumes a ``standard'' population of single stars with
masses $0.6\leq\mathrm{M}\leq120\mathrm{M_\odot}$ and no binary
companions. There is evidence in the literature that accounting for binary
evolution improves the agreement between observations and synthetic
spectra \citep[e.g.][]{Eldridge2009,Eldridge2011}. In such models, WR stars
can manifest over a wider age range, which boosts the UV flux and the LyC
production of their host galaxy. Our predictions for the cluster production 
rates of ionizing photons may therefore be understimated.


\section{Lyman Continuum Escape}\protect\label{sec:discussion}
In Section \ref{sec:ipm} and Figure \ref{fig:ipm}a, IPM based on the \oiii\ and
\oii\ lines revealed a large area of ionized, optically thin gas.  This area includes the
entire central Region A ($2.3$ kpc in diameter), and extending well beyond
the stellar body of the galaxy to the northwest and
southeast of Region A, reaching $\sim3$ kpc in both directions from the
center.  Clusters 1 and 2 in Region A are the most massive and the brightest
objects in the entire galaxy in both FUV and optical (Table \ref{tab:phot}),
and are separated by a projected distance of $280$ pc.  Outside of Region A,
the brightest object is Cluster 8 in Bar I, which is the third most massive
cluster in the galaxy, and on average as bright as some of the Region A clusters.
These three clusters alone cannot account for all of the observed ionized gas.
In what follows, we examine the contribution of all $26$ clusters to the
ionization structure of the ISM. We note that the stellar mass in all
apertures is between $10^5$ and $10^8$ M$_\odot$, as seen in Table
\ref{tab:cigale}.  This means that our clusters are either a congregation of
clusters or an individual SSC. 

We estimate the fraction of diffuse gas emission in Tol 1247$-$232, which
we define as the fraction of the total nebular flux outside of the apertures
for the $26$ clusters in Table \ref{tab:cigale}.  We use the pixel photometry
in Table \ref{tab:phot} to represent the radiation coming from the clusters
and their immediate vicinity. To obtain the emission-line fluxes within the
apertures, we use the continuum-subtracted images obtained in Section
\ref{sec:mu}. For the total nebular flux of the galaxy in each of the
narrowband filters, we use the same fixed aperture of $r=5.8$ arcsec. The
diffuse emission is then estimated as the fraction of flux outside of the
object apertures.   

For \oii, \hb, \oiii, and \ha, we obtain diffuse fractions of $0.83$, $0.89$,
$0.76$, and $0.83$, respectively.  These fractions are much higher than
typical warm interstellar medium (WIM) fractions for starburst galaxies, which
are around $\sim20$ per cent \citep{Oey2007}.  This is because our definition
of the diffuse radiation differs from conventional WIM analysis, in
particular, that our cluster apertures are defined by the stellar light and
therefore are much smaller, excluding the outer areas of the \hii\ regions
associated with each cluster. Our diffuse radiation fraction of $0.83$ in
\ha\ is consistent with \citet{Ostlin2016}, who model the \ha\ emission in
Tol 1247$-$232 pixel by pixel and estimate the diffuse fraction in a
similar fashion. 

Balancing the budget of intrinsic and observed ionizing photons, one can use
the diffuse radiation fraction, and the modeled $Q\mathrm{(H^0)}$ and
$f_{\mathrm{esc}}$ from Table \ref{tab:cigale} to estimate the global escape fraction of
ionizing photons. From the SED fit to the entire galaxy, in Section
\ref{sec:seds_galaxy} we obtained $Q\mathrm{(H^0)_{ion}}=4.9\times10^{54}$
s${}^{-1}$.  With an average diffuse fraction of $0.83$, this
corresponds to a production rate of diffuse ionizing photons of
$Q\mathrm{(H^0)_{diffuse}}=4.1\times10^{54}$ s${}^{-1}$. 
This is the ionizing photon emission rate that the ionizing sources in the
galaxy must account for, and any excess above this number will escape
into the IGM.
From all $26$ clusters, the rate of ionizing photons, escaping
into the ISM is $\sum{Q\mathrm{(H^0)}_i\times f_{\mathrm{esc}, i}}=5.6\times10^{54}$ s$^{-1}$,
where $Q\mathrm{(H^0)}_i$ and $f_{\mathrm{esc}, i}$ are obtained from Table
\ref{tab:cigale}, using parameter values at minimum $\chi^2_\nu$. This implies
that the radiation leaking from the clusters into the ISM can account for $1.37\times Q\mathrm{(H^0)_{diffuse}}$, and
hence the global escape fraction is $0.27$. 
Using the maximum parameter values from Table \ref{tab:cigale} (positive
errorbars), the $26$ clusters can account for $1.55\times
Q\mathrm{(H^0)_{diffuse}}$, and the global escape fraction is $0.35$.
Using the minimum parameter values from Table \ref{tab:cigale} (negative
errorbars), the $26$ clusters can account for only $0.73\times
Q\mathrm{(H^0)_{diffuse}}$, and hence the global escape fraction is zero. We
further note that taking the SED models at minimum $\chi^2_\nu$ at face value,
Region A (Clusters $1 - 7$) dominates the production of
ionizing photons, and alone accounts for $0.83\times
Q\mathrm{(H^0)_{diffuse}}$.

The above estimates for the global $f_{\rm esc}$ are obtained
by ignoring dust absorption inside the \hii\ regions. From the SED modeling of the entire galaxy in Section 
\ref{sec:seds_galaxy}, we obtained a model value for the fraction of ionizing
photons absorbed by dust, $f_{\mathrm dust}$.  The global LyC escape fraction
can then be expressed as,
\begin{equation}\protect\label{eq:one}
  f_\mathrm{esc}^\mathrm{galaxy}  =1 - \frac{Q\mathrm{(H^0)_{obs}}
  }{(1-f_\mathrm{{dust}})\times \sum_{i=1}^{26}{Q\mathrm{(H^0)}_{i}} }
  \quad ,
  \end{equation}
for the case that
$\sum_{i=1}^{26}{Q\mathrm{(H^0)}_{i}}>Q\mathrm{(H^0)_{obs}}$, and zero
otherwise.  Here we assume that $f_{\rm dust}$ is a constant effective fraction
of absorbed LyC photons that applies to all individual clusters.
$Q\mathrm{(H^0)_{obs}}$ is obtained from the observed \hb\ luminosity of
the galaxy, assuming case B recombination, and is estimated to be
$Q\mathrm{(H^0)_{obs}}=4.2\times10^{54}$ s$^{-1}$, after correcting for
internal extinction with $E(B-V)=0.11$. $Q\mathrm{(H^0)}_{i}$ is the model intrinsic LyC
production rate from Cluster $i$, obtained from its Cigale
SED fit in Table \ref{tab:cigale}. The dust fraction at minimum $\chi^2_\nu$ is
$f_\mathrm{dust}=0.30$, as shown in Table \ref{tab:cigale2}.  With these
values one obtains $\sum Q\mathrm{(H^0)}_{i}=6.78\times10^{54}$
s$^{-1}$. From equation \ref{eq:one} the resulting global escape fraction is 
$f_\mathrm{esc}^\mathrm{galaxy}=0.12^{+0.31}_{-0.12}$, which
represents the total isotropic escape in all directions, after accounting for
dust both in the ISM and inside of the HII regions. The uncertainties on
this value are the propagated uncertainties in $f_{\mathrm dust}$ and
$\sum Q\mathrm{(H^0)}_{i}$ from tables \ref{tab:cigale} and \ref{tab:cigale2},
added in quadrature.

We note that this estimate is sensitive to the  
individual escape fractions from all clusters, the estimate of the diffuse  
radiation fraction, the dust absorption fraction, and the modeled intrinsic number of ionizing  
photons.  For example, if the diffuse radiation fraction is $\ge20\% $
lower, then the minimum $Q\mathrm{(H^0)}_i$ can still account for all diffuse  
radiation, and result in a non-zero global escape fraction of $\gtrsim2\%$.  Further,  
SED models of ages younger than $2$ Myr are unavailable in  
Cigale, and hence we cannot model clusters dominated by extremely young, very  
massive stars (cf.  Section \ref{sec:discuss2}) of $\lesssim 1$ Myr.  VMS  
would significantly boost the intrinsic production of $Q\mathrm{(H^0)}_i$ and
further boost the escape fraction.

\begin{figure*}
  \plotone{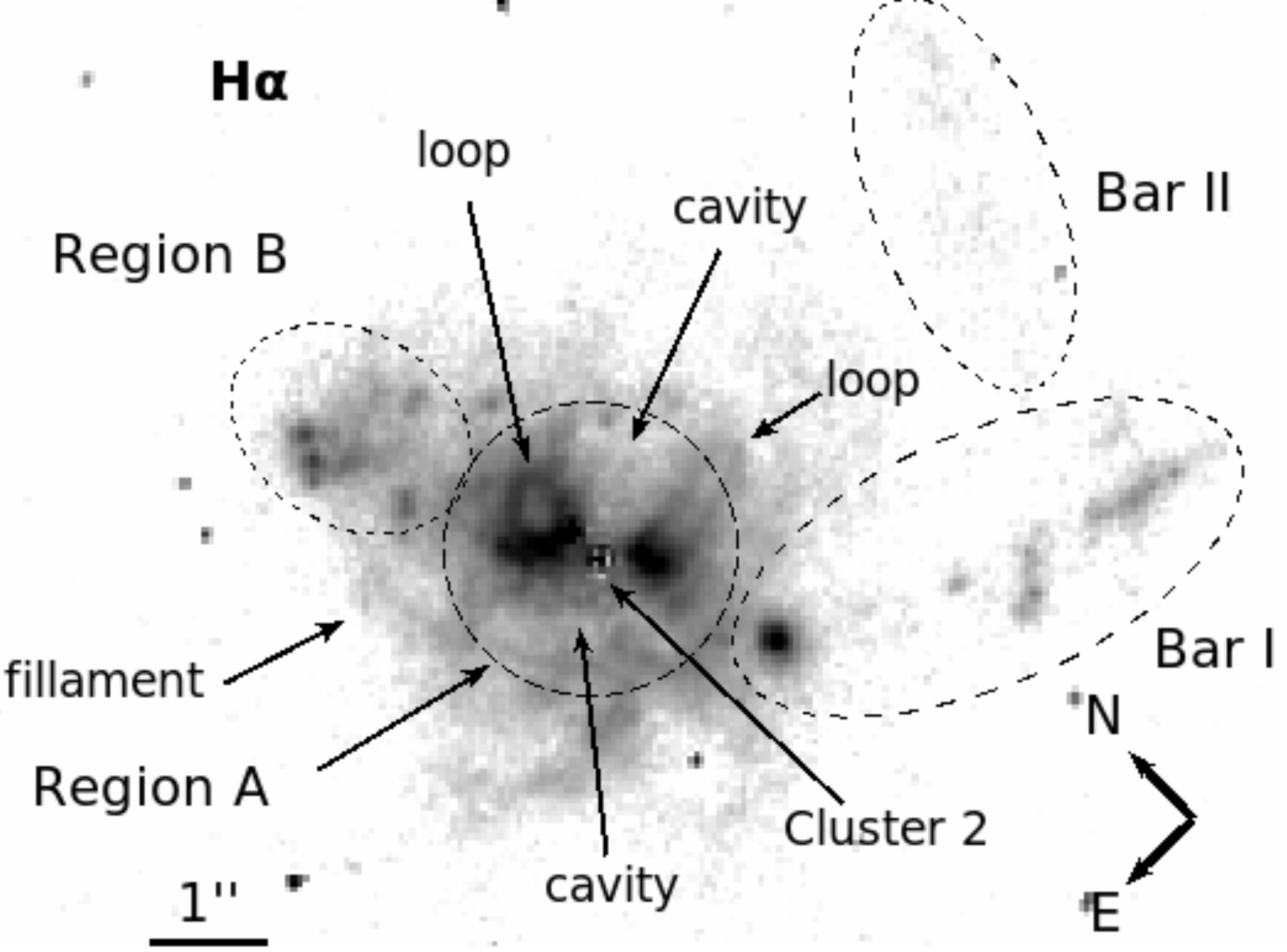}
\caption{Continuum subtracted \ha\ image, showing the structure of the
  ionized gas.  Several loops and filaments are visible in emission in
  and around Region A.  Regions A, B, Bar I and II are indicated with
  dashed ellipses for orientation.
  \protect\label{fig:loops}}   
\end{figure*}

\section{Discussion}\protect\label{sec:discuss2}

The observed LyC escape fraction from Tol 1247$-$232 has
been measured several times through 
direct observations in the LyC regime.  It was first detected with $f_\mathrm{esc} =
0.024_{-0.008}^{+0.009}$ from {\it FUSE} data \citep{Leitet2013}.  The detection was later
confirmed with {\it HST}/COS data but measured to be $f_\mathrm{esc} =0.045\pm 0.012$
\citep{Leitherer2016}.  \citet{Puschnig2017} found a negative flux issue with
the COS reduction pipeline and re-measured $f_\mathrm{esc}$ at $0.015\pm0.005$.  In a
third re-analysis of the COS data, \citet{Chisholm2017} claim that the dark
current has been significantly understimated by \citet{Leitherer2016} and
instead obtain $f_\mathrm{esc}=0.004\pm0.002$.  These same authors predict a higher
$f_\mathrm{esc}$ of $0.05$ from \hi\ absorption properties in their most recent work
\citep{Chisholm2018}, which is consistent with the \citet{Leitherer2016}
measurement.

Our global $f_\mathrm{esc}^\mathrm{galaxy}=0.12^{+0.31}_{-0.12}$ is higher
than these observed measurements, but within the uncertainties, it is
also consistent with zero LyC escape, and is therefore in agreement with
these previous studies. However, taking our estimate at face value,
it is substantially higher than the observed values of $f_{\rm esc}$.
Since the latter are measured in the line of sight, this would suggest that
the escape of ionizing radiation is not isotropic, and would depend on
viewing angle. This is consistent with the 
non-isotropic nature of galactic winds and outflows
\citep{Veilleux2005}. As described earlier, such
mechanical feedback may facilitate LyC escape
\citep[e.g.,][]{Zastrow2011, Zastrow2013}.
Non-isotropic escape via ionized ``tails'', reaching the outskirts of
  the galaxy, has also been suggested for the low-metallicity star-forming
  galaxy SBS 0335-52E \citep{Herenz2017}. 

The highly disturbed and irregular morphology of Tol 1247$-$232, seen in Figure
\ref{fig:rgb}a, suggests a major merger event, typical of starbursts
that are candidate LCEs.  Our SED
analysis indicates the presence of at least two young populations of $\leq4$
and $\sim12$ Myr age (Table~\ref{tab:cigale}), likely the product
of star formation triggered by the 
merger.  The SED modeling also indicates that clusters in Region B and Bar I appear to be
much dustier than those in the central Region A, with average $E(B-V)=0.32$, $0.18$,
and $0.12$, respectively, as seen in Table \ref{tab:cigale}.  Since one does
not expect the average dust attenuation to increase towards the outskirts of a
galaxy, the high $E(B-V)$ values in Region B and Bar I suggest that they may
be remnants of the main bodies of the progenitor merging galaxies.  

In an interacting merger system, mechanical feedback from intense star
formation is expected to play
a significant role in sculpting the morphological structure of the
ISM.  Evidence of mechanical feedback can be observed in the morphology of the
ionized gas, traced by nebular emission.  The continuum-subtracted \ha\ line
image in Figure \ref{fig:loops} highlights numerous loops, filaments and
cavities.  These structures have scales on the order of $\sim 1$ kpc, and
therefore require multiple episodes of star formation \citep[e.g.,][]{Chu2008}.
The two loops and the northwest cavity, indicated in the 
figure, could be multi-supernova superbubbles.  Note the 
symmetric geometry centered on Cluster 2, comprising the two cavities
directly above and below the cluster.  In this projection, the cavities are
perpendicular to the galaxy axis, bisecting Region B and the central
Region A.  These structures apparently correspond to
ionized gas, outlined by filamentary strands of nebular
emission, and lack any substantial stellar component.

We can compare the stellar population and morphology of optically thin regions in
Tol 1247$-$232 with what is seen in other starburst galaxies.  The IPM studies of
NGC 5253 and NGC 3125 revealed narrow ionization cones, most likely powered by clusters
between $1\mbox{-}5$ Myr of age \citep{Zastrow2011, Zastrow2013}.  These
works point out the presence of older stellar populations with ages
$10\mbox{-}100$ Myr from prior star-formation episodes in both galaxies, and
suggest that the ionization cones formed through low density channels
pre-cleared by the older clusters.  This is supported by the
apparently preferred orientation of the ionization cones perpendicular
to the major axis of these galaxies \citep{Zastrow2013}.
The age distribution of the clusters in Tol 1247$-$232 similarly 
indicate a two-stage starburst (Table \ref{tab:cigale} ), where, in addition to the
young objects $\lesssim4$ Myr old, an older population
from a previous star formation episode is also present, with average age
$\sim12$ Myr.  The most prominent cluster from this older population
is the centrally postioned Cluster $2$, which is the most massive
object in the entire galaxy, and rivals its neighbor, Cluster $1$, in
brightness, both in the FUV and optical.  However, 
instead of narrow ionization cones,  Tol 1247$-$232 shows a large,
highly extended area of ionized gas (Figure \ref{fig:ipm}a) centered
on Region A, which reaches the outskirts of the
galaxy to the north-west and south-east.  The presence of 
large-scale, optically thin regions revealed by IPM is consistent
with the known LyC emission from this galaxy and confirms the 
use of this technique in clarifying LyC radiative transfer.
Another LCE showing similar, extended, optically thin morphology through IPM is
Haro 11 \citep{Keenan2017}, where a broad, optically thin region
extends $>1$ kpc from the center of Knot A into the outskirts of the galaxy.
Interestingly, the star formation history in Haro 11 is also
consistent with a two-stage starburst, with a young population of
SSCs having ages $\sim3.5$ Myr, and an older population having
ages $\gtrsim40$ Myr \citep{Adamo2010}. 

Thus, a common feature emerging among galaxies with large-scale
optically thin regions with likely LyC escape, is the two-stage
starburst, in which the episodes of star-formation are separated by $\sim5\mbox{--}40$ Myr.
While narrow ionization cones are seen in the candidate
LCEs NGC 5253 and NGC 3125 \citep{Zastrow2013}, the
two confirmed LCEs Haro 11 and Tol 1247$-$232
reveal much more extensive optically thin ISM.  It is possible that
the morphologies appear different simply due to projection effects,
and IPM of larger samples of LCEs are needed to quantitatively
characterize the ionization structure in these objects.

We note that X-ray emission from an accreting point source has been detected
in both Haro 11 \citep{Prestwich2015} and in region A of Tol 1247-232
\citep{RosaGonzalez2009,Kaaret2017}, which may contribute to the LyC
escape, and help explain the extremely high ionization parameter of $\log
U=-1$, preferred for this galaxy by population synthesis models. 

\section{Conclusions}\protect\label{sec:conclude}
We have used FUV and optical $HST$ imaging of Tol 1247$-$232 to study the
ionization structure of this confirmed LCE via the
technique of ionization-parameter mapping.  The continuum emission in the
\oii$\lambda 3727$, \hb, \oiii$\lambda\lambda 4959,5007$, and \ha\
narrowband filters was first subtracted with the 
mode method of \citet{Keenan2017}, and we demonstrated that this method gives
continuum scaling factors consistent with the skewness method of
\citet{Hong2014} and the pixel-to-pixel method used by, e.g.,
\citet{Boker1999}.  IPM using \oiii\ and \oii\ reveals a large,
optically thin area, engulfing the central region and reaching the outskirts of the galaxy, at
$\sim3$ kpc from the center along the minor axis. Thus, IPM unambiguously
confirms the central region as the origin of the LyC photons that escape in
Tol 1247$-$232. 

We identify $26$ SSCs, seven of which are located in the central, brightest
region of the galaxy, and we model their  SEDs with Cigale.  Our results from
minimum $\chi^2$ SED fitting indicate a population of very young ages of $2-4$
Myr for most clusters.  The two brightest Clusters, 1 and 2, are located in
the central region, and are separated by a projected distance of $280$
parsec. The emerging scenario for the escape of LyC in Tol 1247$-$232 appears
to be a two-stage starburst, in which the older Cluster $2$ ($12$ Myr old) has
generated large-scale superbubbles, loops, and filaments via mechanical
feedback. Young clusters ($\leq4$ Myr) from the second star formation episode,
dominated by Cluster 1, have then ionized the surrounding ISM, facilitated by
this pre-clearing of the region. WN stars have been previously detected in
this galaxy, and we highlight the possibility that these may instead be
unevolved ($\lesssim1$ Myr) very massive stars (VMS).  Their confirmed
presence would revise the age of Cluster 1 accordingly. The LyC luminosity in
the central region is so high that large areas appear to be optically thin,
not just the lowest-density cavities. 

Based on the cluster SED models and observed \hb\ emission in Tol 1247$-$232,
we obtain a LyC escape fraction $f_{\rm esc}^{\rm global} =0.12^{+0.31}_{-0.12}$. The central Region A dominates the
ionization. The 26 clusters can fully account for the observed ionized ISM,
and furthermore can leak LyC with a global non-zero escape fraction. Within
the uncertainties, this is consistent with direct measurements of $f_{\rm
  esc}$ on the order of a few percent in the literature, and also with a zero
escape fraction.  Our larger estimated value compared to the measurements
supports the idea that LyC escape is not isotropic, and may depend on viewing angle.

\appendix

\section{Effects of line contamination in continuum filters}\protect\label{sec:linecont}
The underlying assumption in integrated scaling factor methods is that
any emission lines in the continuum filter have a small contribution compared
to the total continuum flux, which may not be a justified
assumption for starburst galaxies \citep[e.g.][]{Krueger1995}. In our case, F336W, F438W,
and F775W do not contain any strong emission lines, while F547M, used for
subtracting the continuum from the \oiii\ narrowband filter FR505N, 
is affected by the presence of strong \oiii\ emission. \citet{Keenan2017} test their
mode method on synthetic data sets, containing pixels with different continuum
and line properties. Their tests reveal that as long as there are some
continuum-dominated pixels, the mode method determines the
true scaling factor. If no such pixels are present, the scaling factor will be
slightly overestimated. In addition, the presence of the \oiii\ line in the continuum
filter will also cause the continuum flux to be overestimated \citep[e.g.,][]{Kennicutt2008}. This can lead to
an oversubtraction of the continuum and consequently to an underestimation of the \oiii\ line
flux. Following \citet{Kennicutt2008}, we estimate that due to the presence of
\oiii\ in the continuum filter, the effective filter transmission at the
wavelength of \oiii\ is lowered by $\sim80\%$, leading to an underestimate of
the \oiii\ flux by a factor of $\sim5$. We do not apply this correction but
note that our \oiii/\oii\ ratios are therefore lower limits.

\section{Comparison of continuum subtraction methods}
\begin{figure}
\figurenum{A1}
  \gridline{
    \fig{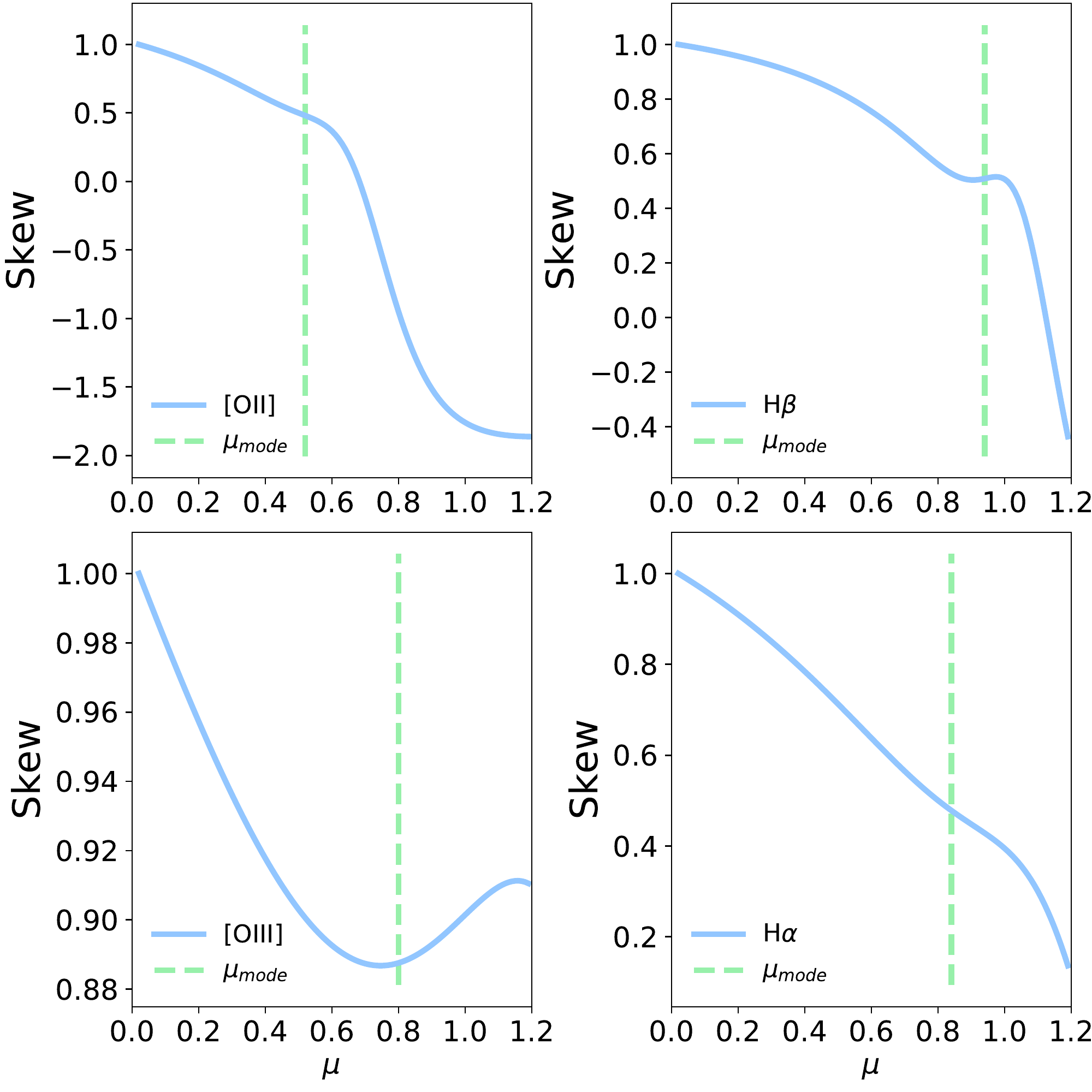}{0.4\textwidth}{(a)}
  }
  \gridline{
    \fig{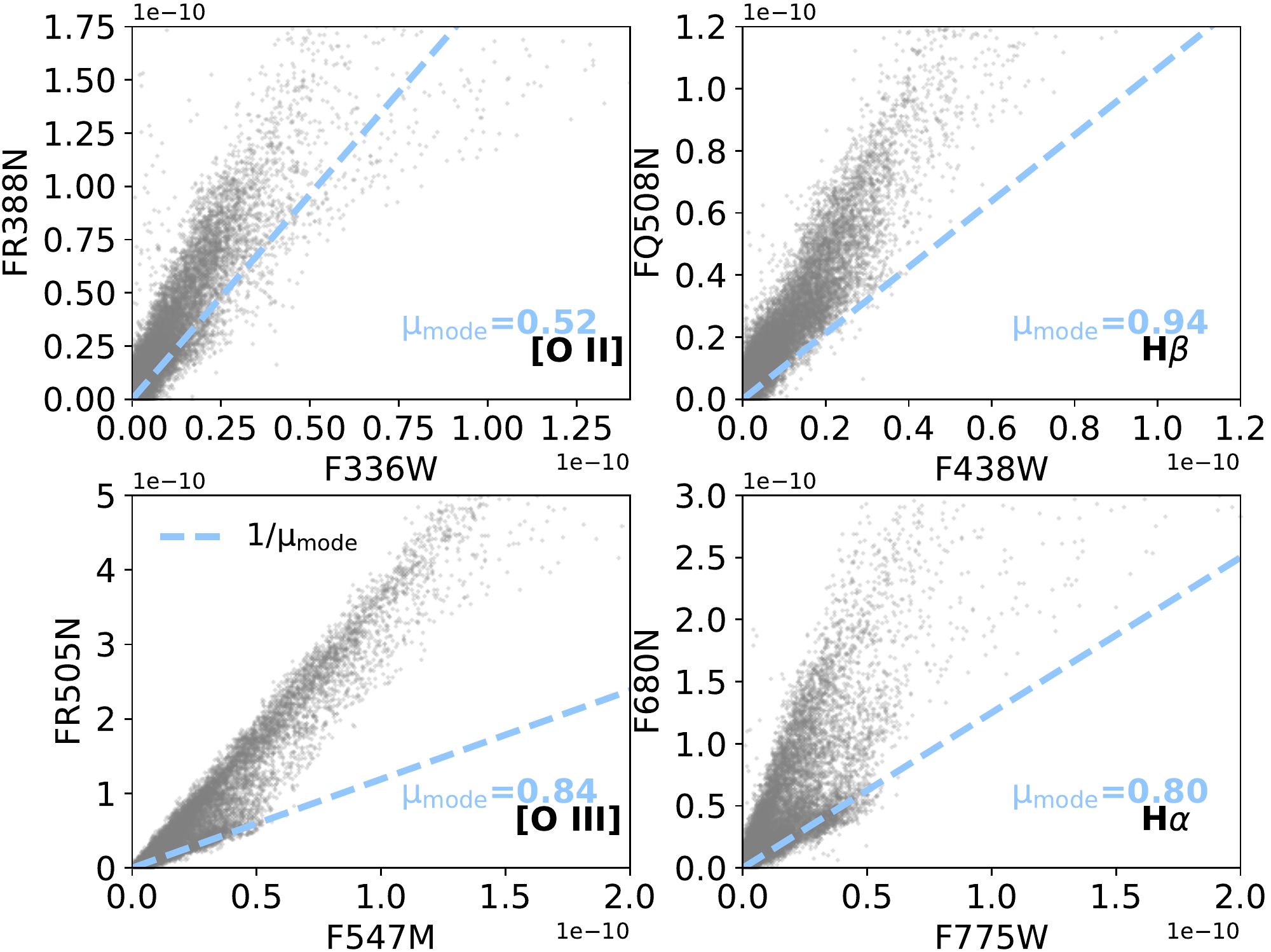}{0.4\textwidth}{(b)}
    }
\caption{(a) \citet{Hong2014} method: Normalized skewness as a
  function of scaling factor $\mu$.  The scaling factor obtained via
  the mode method is indicated by a dashed vertical line.  (b)
  Pixel-to-pixel method: Pixel fluxes in the line filter ($y$-axis) as a
  function of their fluxes in the continuum filter ($x$-axis). The scaling
  factor obtained via the mode method is indicated by a dashed line, and the
  excess values in the narrow-band filters correspond to nebular emission
  above the continuum level. All panels are zoomed in to better show the continuum-dominated
  pixels, found at low flux levels.\protect\label{fig:scalingfactor_appendix}}   
\end{figure}

\protect\label{sec:mu_appendix}
As a sanity check for the mode method of continuum subtraction in Section
\ref{sec:mu}, here we compare with the method from \citet{Hong2014}, shown
in Figure \ref{fig:scalingfactor_appendix}a.  This method uses the
skewness of the pixel flux distribution for the
continuum-subtracted image instead of the mode.  The optimal scaling
factor $\mu$ is again found 
near the transition from undersubtracted to oversubtracted.  In the
case of continuum-dominated pixels, this transition is marked
by a pronounced ``bump'' in the skewness function.  The figure
indicates that the $\mu$ values obtained from the mode method are in
good agreement with the values suggested by the observed
location of the skewness transition bump.

As another check, we also compare with the pixel-to-pixel method
\citep[e.g.,][]{Boker1999,Kennicutt2008} in Figure
\ref{fig:scalingfactor_appendix}b.  This method presents the pixel fluxes in
the line filter as a function of their fluxes in the continuum filter.
In this representation,
continuum-dominated pixels will fall on a straight line, from which a
scaling factor can be recovered as the inverse of the slope.  In
the absence of emission line contribution to the flux, all pixels will
fall on linear relation whose slope depends solely on the relative filter shapes.
As seen by the excess emission in the line filter,
the vast majority of pixels have strong emission-line contributions in
all four line filters (Figure~\ref{fig:scalingfactor_appendix}b).  The
continuum-dominated pixels form the lower, linear envelope
corresponding to the blue dashed line, obtained from
converting $\mathrm{\mu_{mode}}$ from the mode method to a line of the
shown slope. 

\section{Supplementary data from SED fits}
\begin{figure*}
\figurenum{B1}
  \plotone{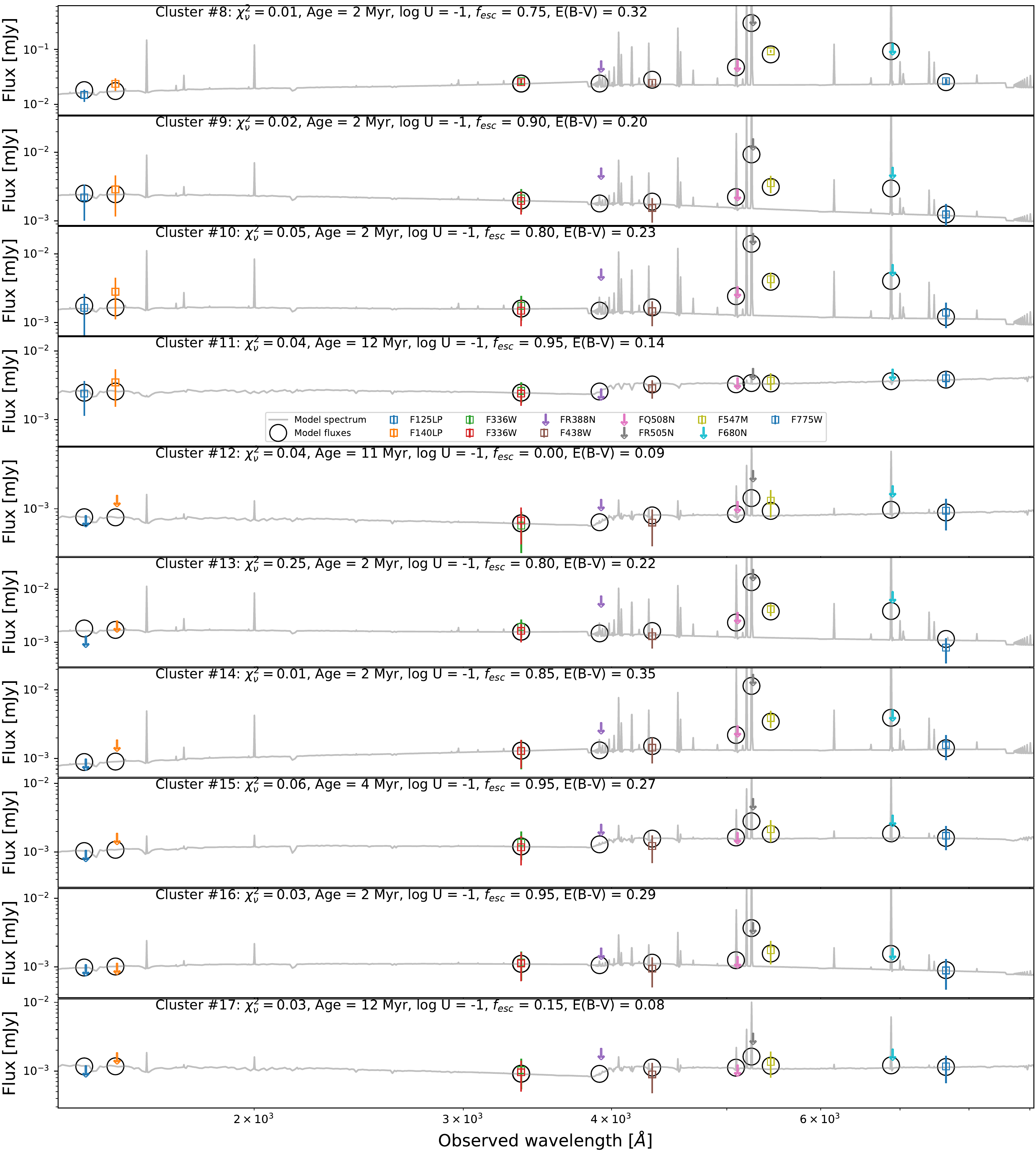}
\caption{Cigale SED fits for Bar I. Photometric data are shown with circles,
  upper limits with arrows. \protect\label{fig:sedbarI}} 
\end{figure*}

\begin{figure*}
\figurenum{B2}
  \plotone{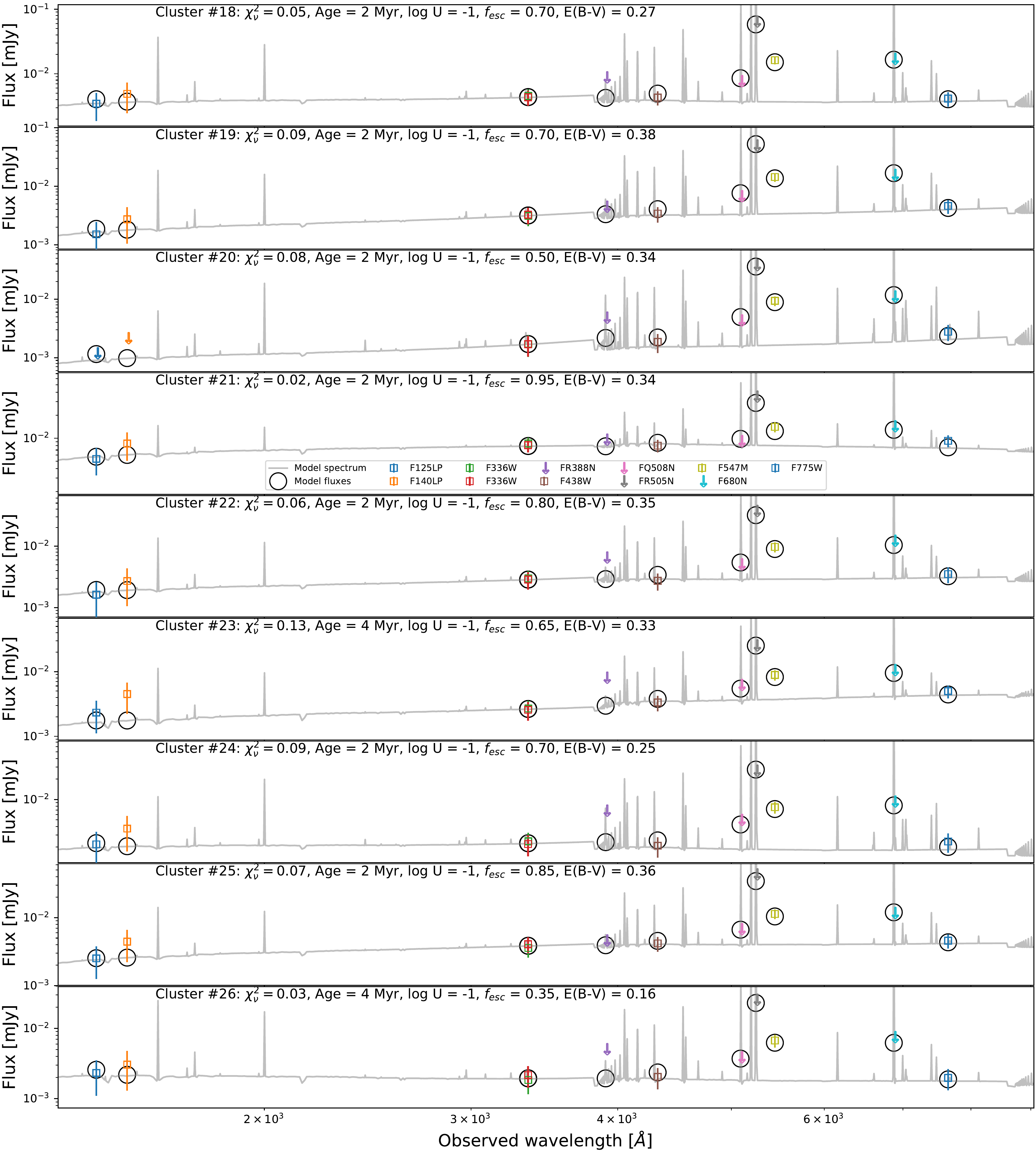}
  \caption{Cigale SED fits for Region B.  Symbols are as in
    Figure~\ref{fig:sedbarI}. \protect\label{fig:sedB}} 
\end{figure*}

\begin{figure*}
\figurenum{B3}
  \gridline{
    \fig{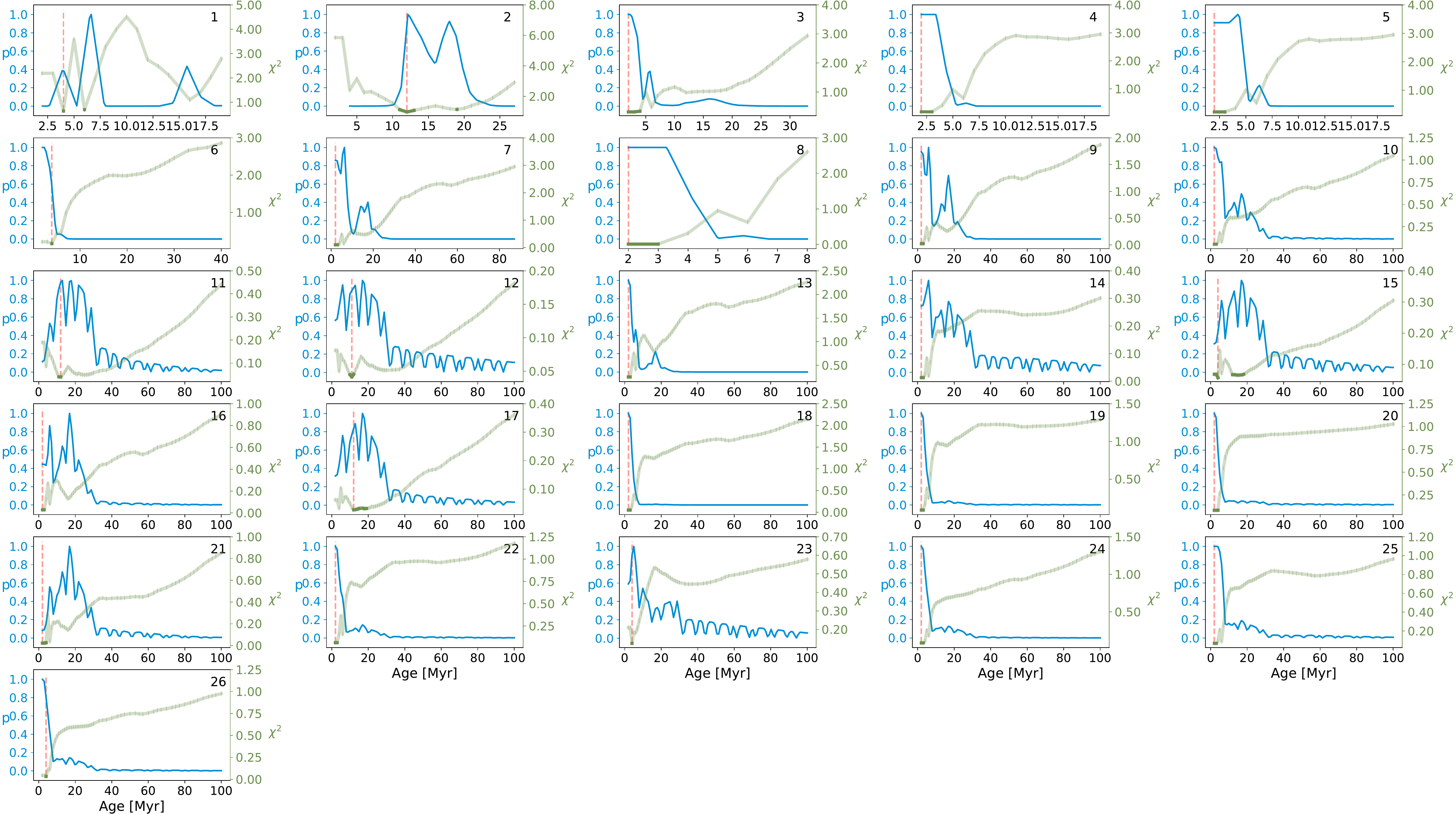}{\textwidth}{}
  }
\caption{PDF of the age parameter for each cluster.  The left $y$-axis
  shows the PDF, color-coded in blue, the right $y$-axis shows the
  corresponding $\chi^2$ distribution, color-coded in green.  The best
  model value, selected by Cigale and corresponding to the lowest
  $\chi^2$, is marked by a dashed red line.  $\chi^2$ values within
  $20\%$ of the minimum represent models indistinguishable from the
  best fit, and are marked by dark green. Note that the used step
  size during the SED fitting can be directly inferred for each parameter from
the density of points in these figures.
  \protect\label{fig:pdf_age}}     
\end{figure*}

\begin{figure*}
\figurenum{B4}
  \gridline{
    \fig{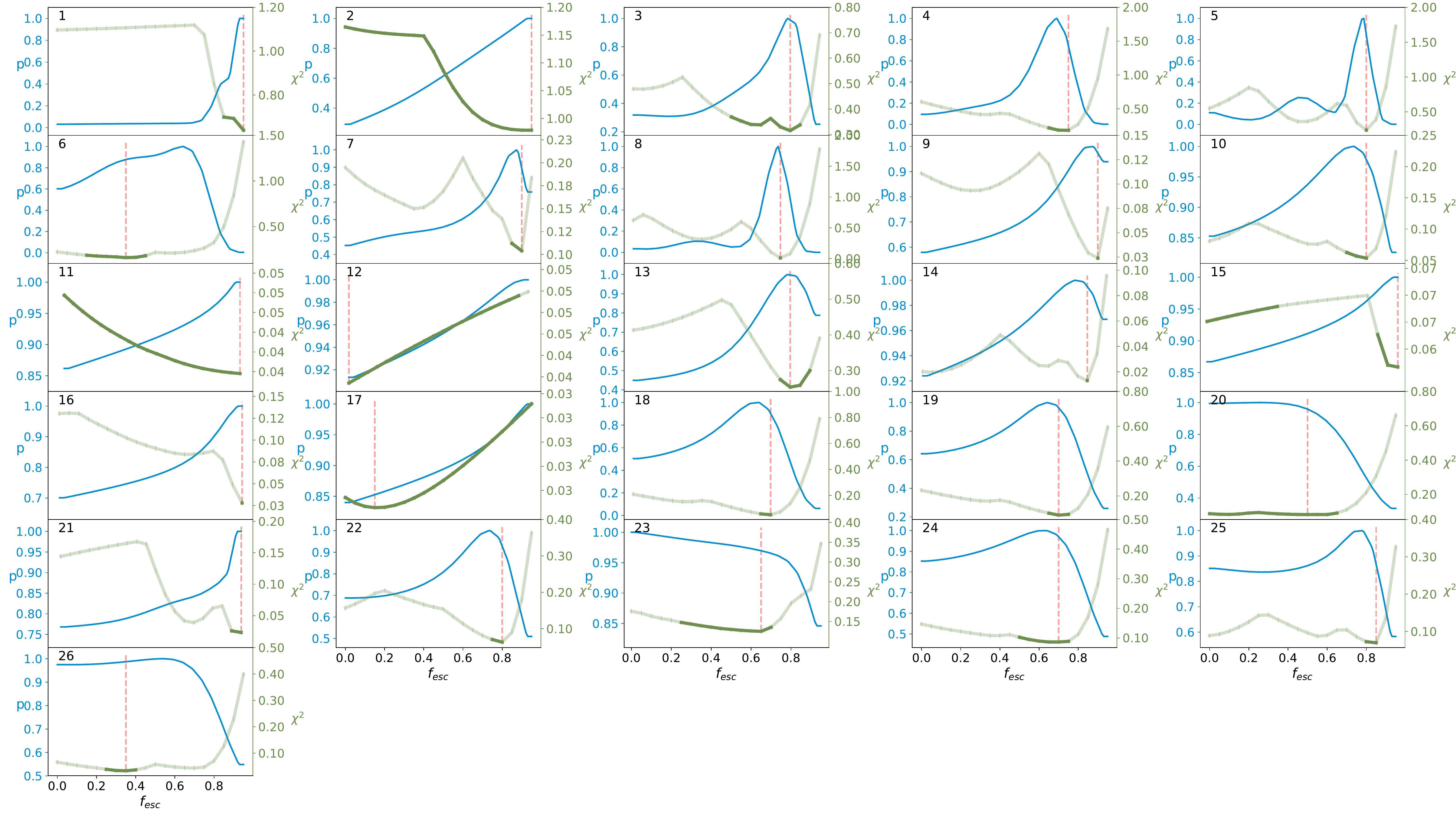}{\textwidth}{}
  }
  \caption{PDF of the $f_\mathrm{esc}$ parameter for each cluster.
Line types and axes are as in
    Figure~\ref{fig:pdf_age}.\protect\label{fig:pdf_fesc}}   
\end{figure*}

\protect\label{sec:seds_pdfs_appendix}
Figures \ref{fig:sedbarI} and \ref{fig:sedB} show the Cigale SED fits for the 
clusters of Bar I and Region B, respectively.
We show, as examples, output model SED parameters, with the probability
density functions (PDF) for age and $f_\mathrm{esc}$ in Figures
\ref{fig:pdf_age} and \ref{fig:pdf_fesc}, respectively.  Light green shows
$\chi^2$, regions with $\chi^2$ within $20$\% of the minimum $\chi^2$ are
indicated by dark green, and the PDF of each parameter with a blue solid
line.

\newpage

\acknowledgments
We thank the referee, Polychronis Papaderos, for the comments and suggestions, which helped
improve this work. The authors are grateful to G\"oran \"Ostlin for helpful discussions and to
M\'ed\'eric Boquien for offering invaluable advice on the Cigale software.  We
also thank Roberto Terlevich for sharing the ground-based spectra of this
galaxy.  This work was supported by NASA grant HST-GO-13702 and the University
of Michigan.



\vspace{5mm}


\bibliographystyle{aasjournal.bst}
\bibliography{tol1247}

\end{document}